\documentclass[twocolumn,showpacs,preprintnumbers,amsmath,amssymb]{revtex4-1}
\usepackage{graphicx}
\usepackage{dcolumn}
\usepackage{bm}

\newcommand{\be}{\begin{eqnarray}}
\newcommand{\ee}{\end{eqnarray}}
\newcommand{\nn}{\nonumber\\}
\newcommand{\la}{\langle}
\newcommand{\ra}{\rangle}

\newcommand{\A}{{\rm A}}
\newcommand{\B}{{\rm B}}
\newcommand{\C}{{\rm C}}

\begin{document}


\title{Real time dynamics and proposal for feasible experiments of lattice gauge-Higgs model simulated by 
cold atoms}

\author{Yoshihito Kuno$^1$, Kenichi Kasamatsu$^2$, Yoshiro Takahashi$^3$, Ikuo Ichinose$^1$, and Tetsuo Matsui$^2$}
\affiliation{
$^1$Department of Applied Physics, Nagoya Institute of Technology, 
Nagoya  466-8555, Japan \\
$^2$Department of Physics, Kinki University, Higashi-Osaka, Osaka 577-8502, Japan\\
$^3$Department of Physics, Graduate School of Science, Kyoto University, Kyoto 606-8502, Japan
}

\date{\today}

\begin{abstract}
Lattice gauge theory has provided a crucial non-perturbative method 
in studying canonical models in high-energy
physics such as quantum chromodynamics.
Among other models of lattice gauge theory, 
the lattice gauge-Higgs model
is a quite important one because it
describes wide variety of phenomena/models related to the Anderson-Higgs mechanism
such as superconductivity,  
the standard model of particle physics, and inflation 
process of the early universe.
In this paper, we first show that atomic description 
of the lattice gauge model allows us to explore real time dynamics 
of the gauge variables by using the Gross-Pitaevskii equations. 
Numerical simulations of the time development of an electric flux
reveal some interesting characteristics of dynamical aspect of the model 
and determine its phase diagram.
Next, to realize a quantum simulator of the U(1) lattice 
gauge-Higgs model on an optical lattice filled by cold atoms, 
we propose two feasible methods: 
(i) Wannier states in the excited bands and
(ii) dipolar atoms in a multilayer optical lattice. 
We pay attentions to respect the constraint of Gauss's law  
and avoid nonlocal gauge interactions. 
\end{abstract}

\maketitle

\section{Introduction} \label{intro}
Cold atoms in an optical lattice have been used as versatile quantum simulators 
for various many-body quantum systems and some important results were
obtained \cite{book}. 
Recently, there appeared several proposals to simulate models of lattice gauge 
theory (LGT) \cite{KK-wilson,KK-kogut,Wiese}.
Since its introduction, LGT has been an indispensable tool 
to studying the non-perturbative aspect
of quantum models in the high-energy physics (HEP) such as confinement of quarks,
the spontaneous chiral-symmetry breaking, etc.
Atomic simulations, if realized, shall certainly clarify the
dynamics, i.e, the time evolution of lattice gauge models, which
is far beyond the present theoretical standard.  

At present, the proposals are classified into two approaches.
In the first approach \cite{Zohar2,Tagliacozzo1,Banerjee1,Zohar3,Zohar4,Banerjee2,
Tagliacozzo2},
atoms with spin degrees of freedom are put on links of the optical lattice.
Such a lattice gauge model is called the ``quantum link model" 
or ``gauge magnet" \cite{Horn,Orland,Chandrasekharan}.
Although the Gauss's law (divergence of the electric field
is just the charge density of matter fields) \cite{KK-ks}  
is assured as the conservation of ``angular 
momentum" \cite{Zohar5}, the Hilbert space of this model itself truncates 
the full Hilbert space of the original U(1) LGT studied in HEP.

In the second approach \cite{Zohar1}, 
one considers the Bose-Einstein condensate (BEC) of atoms put on 
each link of the two or three dimensional optical lattice. 
The U(1) phase variable of the complex amplitude of BEC  
plays a role of dynamical gauge field on the links, and its 
density fluctuation corresponds to the electric field,
the conjugate variable of the gauge field \cite{Zohar1,KK-Tewari}.  
Adopting the phase variable of the atomic field as the gauge field
assures us that we deal with a U(1) field as in HEP   
in contrast with the first approach.
However, to keep the Gauss's law and the short-range gauge interactions
simultaneously, a complex
design of the system and also a fine tuning of interaction parameters 
are generally needed in the experimental setups 
\cite{Zohar1,KK-Tewari}. 

Recently, we proposed a perspective to overcome the difficulty 
to respect the local gauge invariance in the second approach \cite{Kasamatsu}.
Namely,  the cold atomic simulator of the LGT {\em without}
exact local gauge invariance due to untuned interaction parameters
(the Gauss's law is not satisfied) 
can be a simulator for a lattice ``gauge-Higgs" (GH)
model \cite{KK-complementarity} {\em with}
exact local gauge invariance, where the unwelcome interactions 
that violate the Gauss's law are viewed as
gauge-invariant couplings of the gauge field to a Higgs field.

Needless to say, the GH model 
is a canonical model of Anderson-Higgs mechanism and plays a very important
role in various fields of modern physics. Its list includes
mass generation in the standard model of HEP,
phase transition and the vortex dynamics \cite{btj} in superconductivity, 
time-evolution of the early universe such as the dynamics of Higgs phase transition
and the related problems of topological defects, uniformity, etc. \cite{inflation1,inflation2}.
Atomic quantum simulation of this model is certainly welcome because
it simulates the real time evolution of the above exciting phenomena. 

In what follows, we consider the second approach to discuss
the atomic quantum simulator of the U(1) lattice GH model 
by using cold atoms in a two-dimensional (2D) optical lattice. 
Our target GH model has a nontrivial phase structure, i.e., 
existence of the phase boundary between confinement and Higgs phases,
and this phase boundary is to be observed by cold-atom experiments. 
In the experiments, each phase could be generally studied through the non-equilibrium 
dynamics of the system, which are detected by e.g., the density distribution of 
the time-of-flight imaging after the system is perturbed. 
As a reference to such experiments, we make numerical 
simulations of the time-dependent Gross-Pitaevskii (GP) equation 
and observe the real-time dynamics of the atomic simulators. 
In particular, we study the dynamical stability of a single electric flux connecting 
two charges with opposite signs, corresponding to a density hump and dip for the 
atomic simulators.
We stress that this dynamical simulation in an interacting atomic system
gives a new theoretical tool for the analysis of lattice gauge models 
far beyond the present standards of the theoretical study on the LGT 
using the ``classical" Monte-Carlo simulations, the strong-coupling expansion, etc. 
The obtained phase boundary is discussed and compared with that of 
the Monte-Carlo simulations.
Next, we propose two realistic experimental setup for the quantum simulators. 
To respect the constraint of Gauss's law and avoid nonlocal gauge interactions, 
it is necessary to tune suitably the intersite density-density interaction of 
the hamiltonian. We give two ideas: (i) Using Wannier states in the excited bands and
(ii) Using dipolar atoms in a multilayer optical lattice, both of which are 
reachable under current experimental techniques. 

The paper is organized as follows. 
In Sec.\ref{GHMD}, we introduce our target hamiltonian of the U(1) lattice GH model 
starting from the extended Bose-Hubbard (BH) model with the intersite density-density 
interaction. The exact correspondence of the atomic system to the LGT has 
been discussed in Ref.~\cite{Kasamatsu}. 
In Sec.~\ref{fluxdyn}, we present the results of the dynamical simulations by means 
of the GP equation, which is obtained under the saddle-point approximation 
of the real-time path integral of 
the two quantum hamiltonians, i.e., the original BH model 
and the target GH model. The obtained dynamical phase diagram is 
compared with the result of the Monte-Carlo simulations. 
We give two proposals for experiments to construct realistic atomic simulators 
for the lattice GH model in Sec.~\ref{implereal}. 
Method A in Sec.~\ref{higiorbit} relies on extended orbits of the Wannier 
functions in excited bands of an optical lattice. Another method B 
utilize the long-range interaction between atoms in different layers of 
2D optical lattices. Both methods may be possible to tune the intersite 
density-density interactions in the 2D BH system for our purpose. 
Section \ref{sums} is devoted to our 
conclusion and an outlook on future direction. 

\section{From Bose-Hubbard model to the gauge-Higgs model}\label{GHMD}
\begin{figure}[t]
\centering
\includegraphics[width=1\linewidth,bb=0 0 254 254]{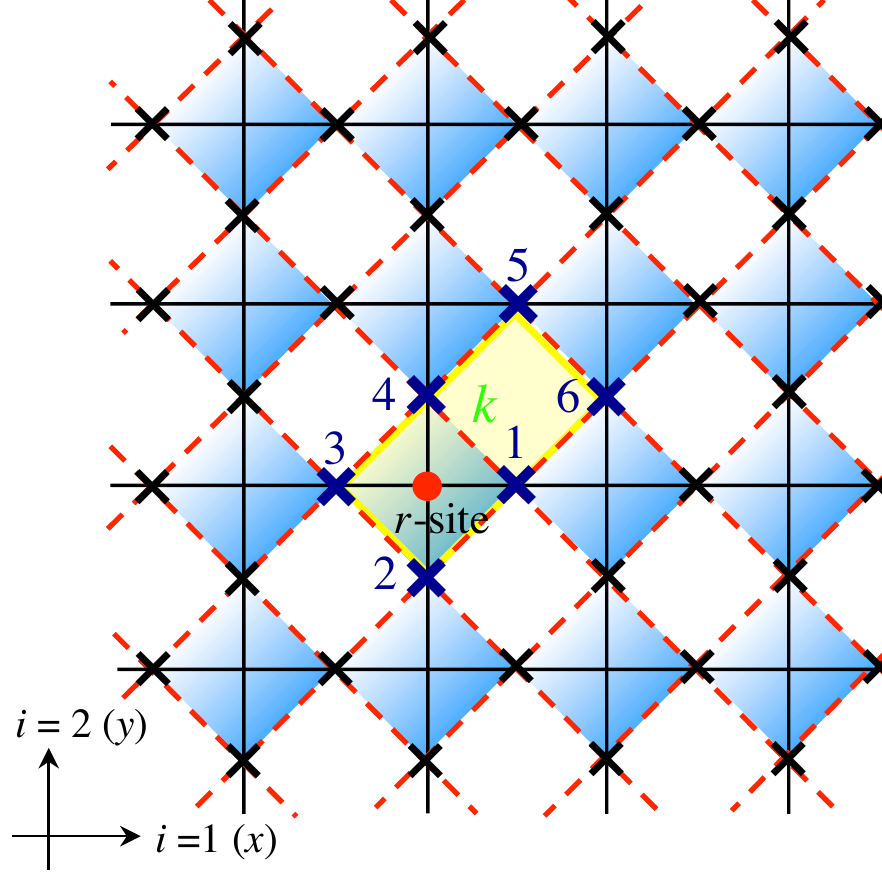}
\caption{Relation between two 2D lattices.
The dashed red lines indicate the 2D optical lattice with the square geometry, 
and cold atoms reside on its sites denoted by black crosses.   
Its unit cell consists of a pair of white and blue squares (yellow region).
The filled black lines indicate the 2D gauge lattice 
on which the U(1) lattice GH model is defined. 
Then the cold atoms are viewed to sit 
on each link of the gauge lattice to play the role of gauge field. 
The relevant sites of the original lattice 
(links among the site $r$ of the gauge lattice) in Table \ref{paratable} 
are numbered as $a=1\sim 6$, as $a=1 = (r,1)$, 
$2=(r-2,2)$, $3=(r-1,1)$, $4=(r,2)$, 
$5=(r+2,1)$, $6=(r+1,2)$, where $(r,i)$ represents the link 
of the gauge lattice emanating from the site $r$ into the 
positive $i\ (=1,2)$-th direction. The gauge lattice plays 
an important role also in the study of the BH model with 
$V_{ab} = 0$ in a different context \cite{mielke}.
}
\label{latticefig}
\end{figure}

Corresponding to the simplest realistic experimental situation 
of the quantum simulator, we focus on the boson system defined on a 2D square lattice. 
We start from a generalized BH hamiltonian \cite{book} 
\begin{align}
\hat{H} = - \sum_{k,a \neq b} J_{ab} \hat{\psi}_{a}^{\dag} \hat{\psi}_{b} 
+  \frac{V_0}{4} \sum_{k,a} \hat{\rho}_{a} (\hat{\rho}_{a}-1)
+ \sum_{k,a \neq b} \frac{V_{ab}}{2} \hat{\rho}_{a} \hat{\rho}_{b}, 
\label{extBHmodel}
\end{align} 
which describes the bosons in a single band of a 2D optical lattice. 
The bosonic atomic fields $\hat{\psi}_{a} = \exp(i \hat{\theta}_{a}) \sqrt{\hat{\rho}_{a}}$ 
are put on the site $a$ of the square optical lattice. The summation is taken over the unit cell 
$k$ (yellow region in Fig. \ref{latticefig}) and, in each unit cell, over the the site $a(b)\in 1$-6. 
We confine ourselves to the nearest-neighbor (NN) and 
next-nearest-neighbor (NNN) couplings for the site pairs $(a,b)$ in the 1st and 3rd terms.
The parameters $J_{ab} (=J_{ba})$, $V_0$, and $V_{ab} (=V_{ba})$ 
are the coefficients of the hopping, the on-site interaction, and the intersite
interaction, respectively, and calculable by using the Wannier 
functions in a certain band. 
The intersite terms $V_{ab}$ may arise when the atoms have a 
long-range dipole-dipole interaction (DDI) \cite{dipolerev}, 
or when the atoms are populated in the excited bands of the optical 
lattice \cite{Scarola}. 

\begin{table}[b]
\centering
\caption{\label{paratable} Atomic parameters $J_{ab}$ and $V_{ab}$ 
in Eq.~(\ref{extBHmodel}). Those not shown below are set to zero to avoid
double counting [$(1,6)$, etc.] or due to longer-ranges [$(3,5)$, etc].}
\renewcommand{\arraystretch}{1.25}
\begin{tabular}{ccp{1.6cm}cc}\hline
group&range&$(a,b)$& $J_{ab}$& $V_{ab}$\\ \hline
(i)&NN &(1,2), (2,3), (3,4), (1,4)& $J$&  
$\gamma^{-2}$\\ \hline
(ii)&1st half of NNN &(1,3), (2,4)& $J'$&  
$\gamma^{-2}$\\ \hline 
(iii)&2nd half of NNN &(1,5), (4,6)& $J''$ &0 \\ \hline
\label{vab}
\end{tabular}\\
\end{table}
To map the BH model onto the hamiltonian of LGT, we consider the diagonal lattice 
whose sites $r=(r_1,r_2)$ are positioned on the centers of the colored squares in 
Fig.~\ref{latticefig}. Then, the original sites can be viewed as links of the diagonal lattice.  
The links are labeled as $(r,i)$ with the direction index $i=1,2$.
To derive the hamiltonian of the target GH model, we consider the case such that 
$J_{ab}$ and $V_{ab}$ take values according
to the following three groups (i)-(iii) for pairs $(a,b)$ of sites as shown 
in Table~\ref{paratable} \cite{Zohar1,Kasamatsu}. 
We note that Table~\ref{paratable} breaks 
the translational symmetry of atomic interactions, e.g., $V_{24}\neq 0$ 
while $V_{15}=0$. Next, we assume that the equilibrium atomic density 
is uniform and sufficiently large 
$\rho_0 \equiv \langle \hat{\rho}_{r,a} \rangle \gg 1$. 
Then, we expand the density operator 
as $\hat{\rho}_{r,i} = \rho_0 + \hat{\eta}_{r,i}$,
and keep terms  up to $O(\hat{\eta}^2)$ to obtain
\begin{align}
\hat{H} &\simeq 
 \frac{1}{2\gamma^2} \sum_{r} \left[ \sum_i  (\hat{\eta}_{r,i} + \hat{\eta}_{r-i,i}) \right]^2 
 + \frac{V_0'}{2} \sum_{r,i} \hat{\eta}_{r,i}^2 \nonumber \\
& -  \rho_0 J \sum_{r,i,\delta} \cos (\hat{\theta}_{r,i} -\hat{\theta}_{r,\delta}) 
 -2 \rho_0 J' \sum_{r,i} \cos (\hat{\theta}_{r,i} -\hat{\theta}_{r-i,i}) \nonumber \\
& -2 \rho_0 J'' \sum_{r,i} \cos (\hat{\theta}_{r,i} -\hat{\theta}_{r-\bar{i},i}),
\label{GHatomhamil}
\end{align}
where $V'_0 \equiv V_0 -2\gamma^{-2} > 0$, $(r,\delta)$ represents the NN links 
of $(r,i)$, and $\bar{1} \equiv 2$, $\bar{2} \equiv 1$. 
The first order term $O(\hat{\eta})$ is absent
due to the stability condition for $\rho_0 = [\mu+4J+2(J'+J'')]/(V_0'+8\gamma^{-2})$ 
with the chemical potential $\mu$. 
In the atomic simulators of LGT, the phase $\hat{\theta}_{r,i}$ plays a role 
of a gauge variable on the link $(r,i)$ and 
its conjugate momentum $\hat{\eta}_{r,i}$ is the electric 
field $-\hat{E}_{r,i}$ \cite{Zohar1,KK-Tewari,Kasamatsu}. 
By replacing $\hat{\eta}_{r,i} \to (-)^r \hat{\eta}_{r,i}$ and 
$\hat{\theta}_{r,i} \to (-)^r \hat{\theta}_{r,i}$ with $(-)^r \equiv (-)^{r_1+r_2}$, 
the first term in the rhs of Eq.~(\ref{GHatomhamil}) 
describes the ``Gauss's law" as
$(2\gamma^2)^{-1} \sum_r [\sum_i  (\hat{\eta}_{r,i} - \hat{\eta}_{r-i,i})]^2 \simeq (2\gamma^2)^{-1} \sum_r (\nabla \cdot \mathbf{E})^2$.  
The two conditions $V_{\rm (i)}=V_{\rm (ii)}
(=\gamma^{-2})$ and $V_{\rm (iii)}=0$ in Table~\ref{paratable} are necessary to generate 
the $(\nabla \cdot \mathbf{E})^2$ term without nonlocal interaction among $E_{r,i}$. 
If these conditions are not fulfilled,   
a product $\hat{E}_{r,i}\hat{E}_{r',i'}$ over the different links
appears additionally, 
and it gives rise to long-range interactions among the gauge field 
$\theta_{r,i}$ in the target GH model. Although such a model still respects gauge symmetry,
we reject it here because all LGTs relevant to HEP are 
generally models with local-interaction.

In Ref.~\cite{Kasamatsu}, it was shown that the partition function 
$Z ={\rm Tr}\exp(-\beta \hat{H})$ 
of the atomic model of Eq.~(\ref{GHatomhamil}) is equivalent to that of the 
GH model. The GH model is the U(1) lattice gauge model on the (2+1)D  lattice, 
and its partition function is given by 
\begin{align}
\hspace{-0.35cm}
Z_{\textrm{GH}}&= \int [dU][d\phi] \exp (A_{\rm I}+A_{\rm P}+A_{\rm L}+
A_{\rm 2I}+A_{\rm I^2}), \nn
\hspace{-0.5cm}
A_{\rm I} &= \frac{1}{2} \sum_{x,\mu}c_{1\mu}(\bar{\phi}_{x+\mu} U_{x,\mu} \phi_x + \mathrm{c.c.}),\nn
\hspace{-0.5cm}
A_{\rm P} &= \frac{1}{2} \sum_{x,\mu<\nu}c_{2\mu\nu} (\bar{U}_{x,\nu} 
\bar{U}_{x+\nu,\mu} U_{x+\mu,\nu} U_{x ,\mu} + \mathrm{c.c.}) ,\nn
\hspace{-0.5cm}
A_{\rm L} &= \frac{c_3}{2} \sum_{x}\Big[
\bar{\phi}_{x+1+2} U_{x+1,2} U_{x,1} \phi_{x}\nn
\hspace{-0.5cm} &+\bar{\phi}_{x+2} \bar{U}_{x+2,1} U_{x+1,2}\phi_{x+1}
+\bar{\phi}_{x} \bar{U}_{x,2} \bar{U}_{x+2,1} \phi_{x+1+2}\nn
\hspace{-0.5cm} &
+\bar{\phi}_{x+1} U_{x,1} \bar{U}_{x,2} \phi_{x+2}
+{\rm c.c.} \Big],\nn
\hspace{-0.5cm}
A_{\rm 2I} &= \frac{c_4}{2} \sum_{x}\sum_{i=1,2}\Big[
\bar{\phi}_{x+i} U_{x,i} U_{x-i,i} \phi_{x-i} 
+ \mathrm{c.c.}\Big],\nn
\hspace{-0.5cm}
A_{\rm I^2} &= \frac{c_5}{2} \sum_{x} \Big[
\bar{\phi}_{x+2+1}U_{x+2,1}\phi_{x+2}\cdot\bar{\phi}_{x+1}U_{x,1} 
\phi_{x}\nn
\hspace{-0.5cm}&+(1\leftrightarrow 2)+ \mathrm{c.c.}\Big].
\label{KK-zgh}
\end{align}
Here, $x=(x_0,r_1,r_2) $ is the site index of the (2+1)D lattice with 
the discrete imaginary time $\tau =x_0 \times \Delta \tau\ [x_0=0 ,\cdots,N_0, N_0
 \Delta \tau=
\beta\equiv (k_{\rm B}T)^{-1}]$ and the 2D spatial coordinate $r_1,r_2$. 
$\mu$ and $\nu$ $(=0,1,2)$ are direction indices. 
The U(1) gauge variables $U_{x,\mu} \equiv \exp(i \theta_{x,\mu})$ 
are defined on the link 
$(x,\mu)$. 
$\theta_{x,\mu}\ (\mu=1,2)$ corresponds to  the eigenvalue of the phase of atomic operator $\hat{\psi}_{r,a}$ through
$ \theta_{x, \mu}= \hat{\theta}_{r,i}$.  
The complex field $\phi_x$ defined on site $x$ is 
a bosonic matter field, referred to ``Higgs field" in the London limit, 
taking the form $\phi_x=\exp(i\varphi_x)$ 
with frozen radial fluctuations. The integration  
$\int [dU] [d\phi]$ is over the angles $\theta_{x,\mu}, \varphi_x \in [0,2\pi)$.
The coefficients $c_1\sim c_5$ are real dimensionless parameters 
for interactions among gauge fields. 
Each term of the action, hence the action itself, and the integration
measure are invariant
under the local U(1) gauge  transformation,
$\theta_{x,\mu}\to \theta_{x,\mu}+\lambda_{x+\mu}-\lambda_x,\
\varphi_x\to\varphi_x+\lambda_x$.

According to Ref.~\cite{Kasamatsu}, 
the atomic simulator of the GH model in a 2D system 
corresponds to the following case of parameters for $c_{1\mu}, c_{2\mu\nu}$: 
\begin{align}
&c_{10}=c_1, \hspace{5mm} c_{11}=c_{12}=0,\nn
&c_{201}=c_{202}=c_2, \hspace{5mm} c_{212}=0.
\label{atomicparam}
\end{align}
In terms of the atomic system, the $c_1$- and $c_2$-terms
describe the sum of the self coupling and  
the neighboring correlations of densities of atoms,
and the $c_3$- and $c_{4,5}$-terms describe 
the NN and the NNN hopping terms respectively. 
The relations among the parameters of 
Eq.~(\ref{GHatomhamil}) and Eq.~(\ref{KK-zgh}) are
\begin{align}
&c_{1}=\frac{\gamma^2}{\Delta\tau},\ \ \ \
c_{2} = \frac{1}{\Delta\tau V_0'},\ \ \ \
c_{3} = 2 J \rho_0 \Delta\tau,\nn
&c_{4} = 2 J' \rho_0 \Delta\tau,\ \ \ \
c_{5} = 2 J'' \rho_0 \Delta\tau, 
\label{parameterrelation}
\end{align}
In experiments, we expect low $T ( \lesssim 10$ nK set by the parameters of $\hat{H}$), and  
the quantum phase transitions may be explored in a multi-dimensional space parametrized
by the dimensionless and $\Delta \tau$-independent combinations such as
$c_1/c_2=\gamma^2V_0', c_3 c_2=2J\rho_0/V_0',$ etc.

\section{Real-time dynamics of simulators: stability of an electric flux} 
\label{fluxdyn}
\begin{figure*}[t]
\centering
\includegraphics[width=1.0\linewidth,bb=0 0 794 269]{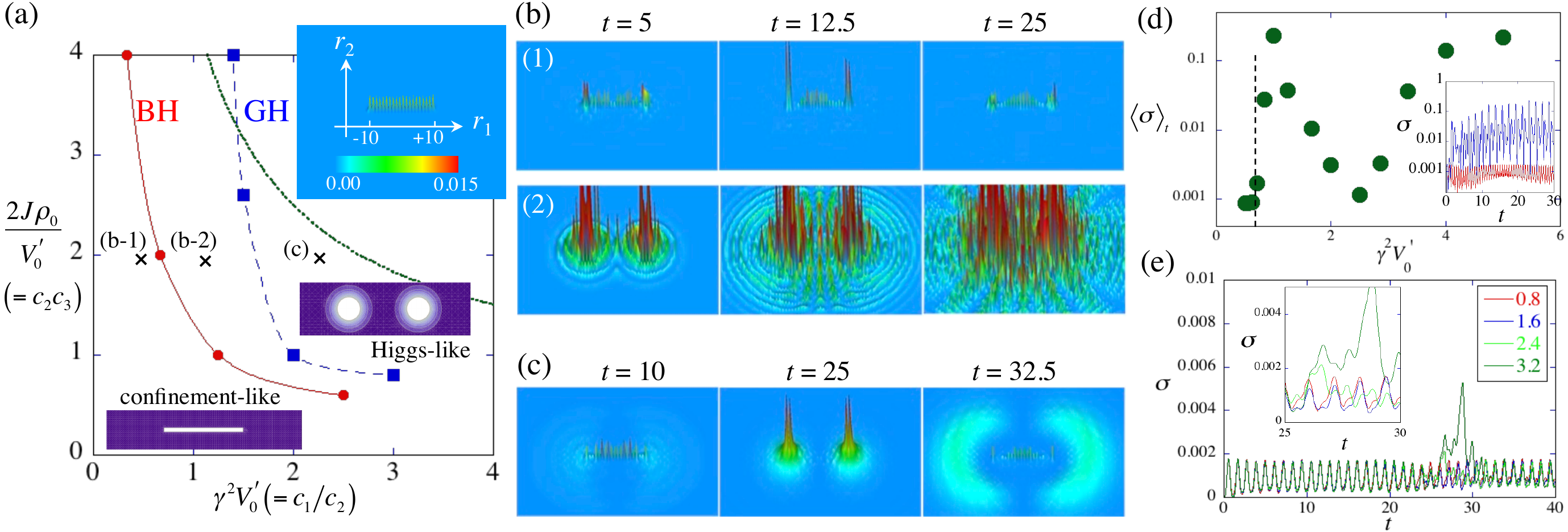}
\caption{Results of the dynamical simulations. 
(a) The dynamical phase diagram with respect to the $\Delta \tau$-independent parameters 
($\gamma^2 V_0'$)-($2J\rho_0/V_0'$) in the BH and GH models. 
In terms of the GH model Eq.~(\ref{KK-zgh}), the horizontal and vertical axes correspond to 
$c_1/c_2$ and $c_3 c_2$, respectively. 
The two models give the different phase boundaries, where 
Higgs (confinement) like behavior can be observed at the right (left) of each boundary. 
The phase boundary of the BH model is calculated for fixed $\rho_0=1$.
The dotted curve gives a $\mu = 0$ curve for $\rho_0=1$, where the amplitude dynamics 
of the BH model is frozen so that the overall dynamics can coincide to that of the GH model. 
The upper right inset shows the initial configuration of the squared density $\eta^2 = (\rho - \rho_0)^2$ 
in the simulation. We put the density modulation $\eta_{r,1} = (-)^{r} 0.1\rho_0$ for $-R \leq r_1 \leq R-1$ 
with $R=10$. 
(b) Time development of the density fluctuation $\eta^2$ for the BH model. 
The parameters are given as $(\gamma^2 V_0',2J/V_0') =$ (0.625,2.0) (b-1, upper panels) 
and $(1.25,2.0)$ (b-2, lower panels), corresponding to the confinement and Higgs regime, 
respectively (the parameters are also shown in (a) by crosses). 
The unit of time is chosen as $\hbar/V_0'$, which is $\sim 0.7$msec 
for the typical energy scale $V_0' \sim 10$nK. 
A flux keeps its initial configuration in the confinement phase (b-1), 
while it is disappeared by the density fluctuation in the Higgs phase (b-2).
(c) The same in (b) for the GH model with the parameter 
$(\gamma^2 V_0' ,2J\rho_0/V_0') =$ (2.4,2.0) of the Higgs regime. 
The large amplitude density wave is generated at the point charges, 
being emitted intermittently. In the confinement phase, the dynamics is 
similar to Fig.~(b-1). 
(d) Plot of the time average of the remnant of flux $\sigma$ of Eq. (\ref{sigma}) 
for the BH model. The shown results are obtained for $\rho_0=1$ and 
$2 J \rho_0 / V_0'=2$ and some values of $\gamma^2 V_0'$. 
The vertical dashed line gives the boundary between the confinement (left) and Higgs (right) regime.
The inset shows 
the time evolution of $\sigma(t)$ for (b-1) (red curve) and (b-2) (blue curve). 
(e) Plot shows the evolution of $\sigma(t)$ for the GH model with $2J\rho_0/V_0'=2$ 
and several values of $\gamma$ in the legend. 
The left inset is the enlarged view during $t \in [25,30]$, where $\sigma$ 
grows intermittently.}
\label{Fluxremgp}
\end{figure*}
In actual experiments observing the nonequilibrium 
time-evolution  of a quantum simulator, the results
globally reflect the phase structure of the target model.   
The (2+1)D GH model supports the confinement phase and the 
Higgs phase (see Appendix~\ref{app1}). 
The confinement phase is characterized by the strong phase fluctuation; 
when static two point charges, such as density defects created by the 
focused potentials, 
are put on, they are connected by an almost straight  electric flux (linearly-rising 
confinement potential). 
In contrast, the Higgs phase possesses the phase coherence 
over the system and the system can be regarded as a superfluid phase; 
the density wave can propagate around the charges \cite{Kasamatsu}. 

To get some insight on the time-evolution of the system, 
we study the dynamical features of the simulators through 
numerical simulations under the mean-field approximation of the 
two quantum hamiltonians: the base BH model Eq.~(\ref{extBHmodel}) 
and the target GH model Eq.~(\ref{GHatomhamil}). 
The time-dependent equations can be derived 
from the real-time path-integral formulation under 
the saddle-point approximation (we put $\hbar=1$).
The operators of the original hamiltonian are replaced by the 
$c$-number fields. 
We confine ourselves to the models with only 
NN hopping $J \neq 0$ and $J' = J'' =0$ for simplicity. 
We note in advance that the mean-field equations necessarily underestimate 
quantum fluctuations, and their results should be taken as a guide 
to practical and future experiments which are expected to reveal the 
real dynamics of quantum systems. 

The equation of motion for $\psi$ in the BH model of Eq.~(\ref{extBHmodel})
can be derived from the Lagrangian 
$L =  - \sum_{r}\sum_{i=1,2} i \psi_{r,i}^{\ast} (d \psi_{r,i}/dt) - H$.
It  is the discretized version 
of the GP equation called the discrete nonlinear Schor\"{o}dinger 
equation \cite{Kevbook} and given by
\begin{align}
i \frac{\partial \psi_{r,i}}{\partial t} = - J ( \psi_{r,\bar{i}} +
\psi_{r-\bar{i},\bar{i}} + \psi_{r+i,\bar{i}}+\psi_{r+i-\bar{i},\bar{i}}) \nonumber \\
+ \biggl[  \left( V_0' + \frac{2}{\gamma^2} \right) 
|\psi_{r,i}|^2 + \frac{1}{\gamma^2} 
(|\psi_{r,\bar{i}}|^2 + |\psi_{r-\bar{i},\bar{i}}|^2   \nonumber \\
+ |\psi_{r+i,\bar{i}}|^2 + |\psi_{r+i-\bar{i},\bar{i}}|^2 
+ |\psi_{r-i,i}|^2 + |\psi_{r+i,i}|^2) \biggr] \psi_{r,i},
\label{DNLSEsim}
\end{align} 
where $i=1,2$ and $\bar{1}\equiv2,\bar{2}\equiv1$.
The uniform stationary solution can be obtained by substituting $\psi_{r,i} = \psi_0 e^{-i \mu t}$ as 
$|\psi_0|^2 = (\mu + 4 J)/(V_{0}' + 8 \gamma^{-2})$, where $\mu$ is the chemical potential. 
Since an important quantity to observe the 
dynamics of electric fluxes is the density fluctuation, we give the 
equilibrium density $|\psi_0|^2 = \rho_0$ by controlling 
the chemical potential as $\mu = \rho_0( V_0' + 8 \gamma^{-2}) -4J$ and 
see the evolution of the density fluctuation $\eta = \rho - \rho_0$. 

The time-dependent equation of motion for $\eta$ and $\theta$ in the GH model of
Eq.~(\ref{GHatomhamil}) is derived in the similar way from  
$L =  - \sum_{r,i}  \eta_{r,i} (d \theta_{r,i}/dt) - H$ as 
\begin{align}
\frac{d \eta_{r,i}}{dt} = &2J\rho_0 \sum_{j} \sin (\theta_{r,i} - \theta_{r,j}), \label{densityligp} \\ 
\frac{d \theta_{r,i}}{dt} = &- V_0' \eta_{r,i}  - \frac{1}{\gamma^2} (\eta_{r,i} + \eta_{r-i,i} + \eta_{r,\bar{i}} + \eta_{r-\bar{i},\bar{i}}) \nonumber \\ 
&- \frac{1}{\gamma^2}  (\eta_{r+i,i} + \eta_{r,i} + \eta_{r+i,\bar{i}} + 
\eta_{r+i-\bar{i},\bar{i}})  .  \label{lgp} 
\end{align}
In terms of the optical lattice, the summation over $j$ 
of Eq.~(\ref{densityligp}) implies to take over the four atomic sites 
which are NN to the atomic site $(r,i)$ ($i=1,2$). 
In terms of the gauge lattice, given an atomic link $(r,i)$,
$(r,j)$ takes $(r,\bar{i}\:), (r-\bar{i},\bar{i}\:), 
(r+i,\bar{i}\:), (r+i-\bar{i},\bar{i}\:)$. 
Equations~(\ref{densityligp}) and (\ref{lgp})  can be also derived by linearizing 
Eq.~(\ref{DNLSEsim})  with respect to the density 
$\rho_{r,i} (t) = \rho_{0} + \eta_{r,i} (t)$. The constraint of the Gauss's law 
requires the replacement $\eta_{r,i} \to (-1)^r \eta_{r,i}$ and $\theta_{r,i} \to (-1)^r \theta_{r,i}$.
We make a dimensionless form of 
Eqs.~(\ref{DNLSEsim})-(\ref{lgp}) by using the energy scale $V_0'$. 
In solving both set of equations of motion, 
we use the $200 \times 200$ discretized space 
and the time step $\Delta t = 10^{-4}$. 

As an explicit example to apply the dynamical equations,
we consider the dynamical stability of a single straight flux connecting two external charges, 
which is prepared as an initial condition. 
In the confinement phase, a set flux string should be stable. 
To see the stability of the flux configuration, we put the density modulation 
$\eta_{r,1} = (-)^{r} 0.1\rho_0$ for $-R \leq r_1 \leq R-1$ in the background 
initial density $\psi_{0} = 1$, in which the length of the flux is $R=10$.
The presence of point charges is taken into account by fixing 
$\eta_{R,1} = 0.1\rho_0$ and $\eta_{R-1,1} = -0.1\rho_0$ through the time evolution. 
The free parameters of this system are $(\gamma^2,V_0',J)$, related to 
$(c_1,c_2,c_3)$. By using the $\Delta \tau$-independent parameters, 
we expect the confinement (Higgs) phase for small (large) values of 
$c_1/c_2=\gamma^{2} V_{0}'$ and $c_3 c_2= 2 J \rho_0/V_{0}'$ (see Appendix~\ref{app1}). 

Figure \ref{Fluxremgp}(b) and (c) represent the time evolution of the density 
distribution $\eta^2$ calculated by the above two models.
For a certain value of $J$, both models show similar behaviors for small values of $\gamma^{2}$, where 
the placed density flux is stable and does not spread out. This captures 
the characteristics of the confinement phase with strong phase fluctuation, 
where the density fluctuation can be localized by the mechanism similar to the 
self-trapping effects as observed in a cold atom experiment \cite{Reinhard}. 
However, the underlying physics is slightly different because the system in 
Ref.\cite{Reinhard} possesses only on-site interaction, without long-range one. 
With increasing $\gamma^{2}$, i.e., 
the Higgs coupling, the structure of the density flux is 
gradually lost by emitting the density waves from the charge. 
This emission is a characteristic of the superfluid phase, i.e, Higgs phase, 
where the phase-coherence can generate a long-wavelength phonon. 
The density waves are generated in a different way: 
successively in the BH model and intermittently in the GH model, 
propagating concentrically around the point charges with the sound velocity 
$\sim \sqrt{J \rho_0 V_0'}$ for $\gamma^2 \gg 1$.

To judge whether the system is in confinement- or Higgs-regime by dynamical simulations, 
we calculate the remnants of the flux $\sigma (t)$ defined by 
\begin{equation}
\sigma (t) = \sum_{\ell \in \mathrm{initial\ flux\ line}} [\eta_\ell^2 (t) - 
\eta_\ell^2 (0) ]^2,
\label{sigma}
\end{equation}
where the sum is taken over the sites on which the density flux line is set 
initially. The flux is stable when $\sigma$ is kept small during the time evolution. 
Figure \ref{Fluxremgp} (a) shows the dynamical phase diagram obtained by 
the behavior of $\sigma$ shown in Fig.~\ref{Fluxremgp}(d) and (e). 
The rapid oscillation of $\sigma$ reflects in the periodic vanish-revival cycle of the density flux. 
In the BH model, we calculate the time-average $\langle \sigma \rangle_t$ and determine the 
phase boundary by finding the point at which $\langle \sigma \rangle_t$ almost vanishes 
(below 0.001; see Fig.~\ref{Fluxremgp}(d)). 
In the GH model, the boundary is determined by the appearance of rapid growth of 
$\sigma(t)$ due to the intermittent density-wave emission as seen Fig.~\ref{Fluxremgp}(e). 

It is important to note that our dynamical approach can give a new method to 
explore the phase structure of the LGT. The validity of our approach exactly 
stems from the correspondence of the 
LGT to the theoretical description of the atomic systems in Sec.~\ref{GHMD}. 
Although the dynamical results are obtained under the mean field approximation and 
only applicable to the GH model with 
the unitary gauge of the Higgs field \cite{Kasamatsu},  
the dynamical phase boundaries of both models are qualitatively 
in good agreement with the result of the Monte Carlo simulations of the full GH model of Eq.~(\ref{KK-zgh}).
(see Fig.~\ref{phasestructure} and Appendix~\ref{app1}). 

The dynamical difference of the BH and GH models can be observed in the amplitude 
fluctuation of the simulating gauge field. 
Because the GH model is obtained by expanding $\rho = \rho_0 + \eta$ around 
the constant density $\rho_0 \gg 1$, the BH model can approximately reproduce the 
GH model when the Thomas-Fermi limit is satisfied; note that 
the boundary of the BH model in Fig.~\ref{Fluxremgp}(a) is obtained for 
the particular value $\rho_0=1$.
In addition, near the situation $\mu=0$ represented by a dotted curve in Fig.~\ref{Fluxremgp}(a), 
the density fluctuation is accidentally frozen because the development of the 
homogeneous wave-function is driven as $\psi_0 e^{-i \mu t}$.
Then, the dynamics of the BH model 
is similar to the GH model. This is a reason of the decrease of $\langle \sigma \rangle_{t}$ 
around $\gamma^2 V_0'=2.5$.
Another point is that the amplitude fluctuation in the BH model 
can give rise to a similar effect of the fluctuation of the Higgs coupling. 
When the Higgs field moves away from the London limit, the Higgs-confinement transition may become 
first order and its boundary can be sharp \cite{Wenzel}. 
Since our GH model corresponds to the London limit, in which the amplitude fluctuation of 
the Higgs field is absent, the phase boundary becomes less clear 
because the two phases connect with each other through crossover. 
The significant amplitude fluctuation in the BH model can lead to 
the stabilization of the Higgs phase as seen in Fig.~\ref{Fluxremgp}(a).

\section{Implementation with cold atoms}\label{implereal}
In this section we present two methods to realize $V_{ab}$ as shown in Table~\ref{paratable}. 
A major way to prepare intersite interactions in BH systems is 
to use DDI between atoms or molecules \cite{dipolerev,Yan,Paz}. 
In usual experiments, dipoles of an atomic cloud are uniformly polarized along a 
certain direction, and one may easily check that uniformly 
oriented dipoles generate $V_{ab}$ {\it different from}
the configuration of $V_{ab}$ in Table~\ref{paratable}. 
This is partly because we consider a square lattice, and 
the similar requirement for $V_{ab}$ is satisfied on the triangular or Kagome 
lattice \cite{KK-Tewari}.
Although an individual control of the polarization of a dipole 
at each site may achieve $V_{ab}$ in Table~\ref{paratable}, 
its actual fulfillment is difficult (some discussions can be seen in the system of 
polar molecules \cite{Gorshkov}), 
and importantly the hopping process between sites with different dipole 
orientations are prohibited or reduced due to the conservation of the atomic spin. 
We note that the bipartite structures of the nanoscale ferromagnetic islands have 
been proposed for realizing the right Gauss law constraint using dipolar interactions \cite{Moessner}.
Recently, there is an interesting proposal to realize 
$V_{ab}$ in Table~\ref{paratable} by using the Rydberg $p$-states of cold atoms \cite{Glaetzle}.

In Sec.~\ref{higiorbit}, we discuss the possibility to realize the values of $V_{ab}$ in Table~\ref{paratable} by
using the excited bands of an optical lattice, which is an alternative route to 
get intersite interactions \cite{Scarola}. 
In Sec.~\ref{kunolayer}, we discuss a system of multi-layer 
2D optical lattices \cite{Macia} to realize 
tunable DDI between of atoms. The difference from the proposal 
in Ref.~\cite{Moessner} is that the long-range interaction of dipoles 
between {\it different} layers is controlled by tuning the height 
of the two layers and the length of dipoles in Ref.\cite{Moessner}, 
while in our case, the long-range interaction in the {\it same} layer 
is controlled through the mediation of atomic interaction in {\it different} layers. 
These proposals are within reach in current experimental techniques.

\subsection{Method A: Using excited bands of an optical lattice}\label{higiorbit}
\begin{figure}[t]
\centering
\includegraphics[width=1.0\linewidth, bb=0 0 623 623]{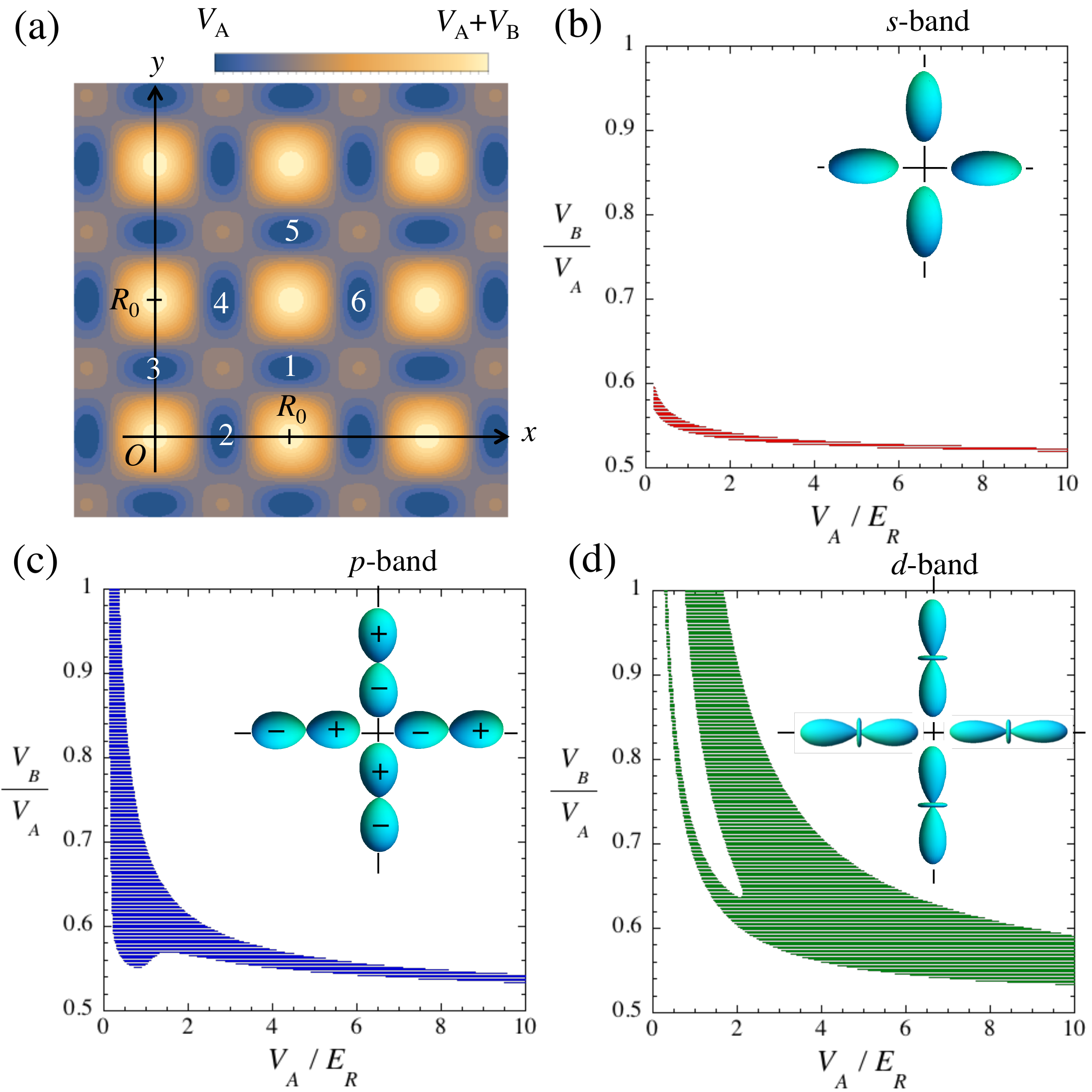}
\caption{Method A to realize the value of $V_{ab}$ in Table~\ref{paratable} using 
excited bands of an optical lattice. (a) The profile of the optical lattice 
of Eq.~(\ref{suitpot}) for $V_\mathrm{A}=V_\mathrm{B}$. 
The minima of the lattice are located at $(x,y) = R_0 (m_x,m_y+1/2)$ for 
the horizontal links and $(x,y) = R_0 (m_x+1/2,m_y)$ for 
the vertical links ($m_{x,y}$: integers), where $R_0 = \pi/k$ is the lattice constant. 
The panels (b)-(d) show the domains (colored region) that satisfy the approximated condition 
for the overlap integral $0.95 < O_{12}/O_{13} <1.05$ 
and $O_{15}/O_{12} \leq 0.1$ in the $V_\mathrm{A}/E_R$-$V_\mathrm{B}/V_\mathrm{A}$ 
plane for $s$-, $p$-, and $d$-orbitals, respectively. 
For $p$-, and $d$-orbitals, the domains bifurcate due to the radial peak structure of the 
Wannier function (see Appendix~\ref{app2}). 
}
\label{bandsetup}
\end{figure}

The Wannier functions in excited bands have extended anisotropic 
orbitals compared with the lowest $s$-orbital band. Thus, we expect the significant 
intersite density-density interaction without introducing DDI between atoms \cite{Scarola}. 
To implement this scheme, we assume the following optical lattice potential:
\begin{align}
V_{\mathrm{OL}} &= V_{\mathrm{A}} \left( \cos^2 kx + \cos^2 ky \right) \nonumber \\
&+ V_{\mathrm{B}} \left[ \cos^2 k(x-y) + \cos^2 k(x+y) \right], \hspace{3mm} V_\mathrm{A}, V_\mathrm{B} \geq 0,
\label{suitpot}
\end{align}
which can be created in a current experimental setup.
For $V_{\mathrm{B}}/V_{\mathrm{A}} > 0.5$, the potential forms a checkerboard lattice 
(line graph of a square lattice \cite{mielke}) and its minima are characterized by 
anisotropic harmonic form as shown in Fig.~\ref{bandsetup}(a). 
This anisotropy is necessary to prevent the intraband mixing dynamics.  
Excitation to higher orbitals can be achieved by stimulated Raman transition \cite{Muller} 
or nonadiabatic control of the optical lattice \cite{Wirth,Olschlager}.

The intersite density-density interaction is proportional to the overlap integral 
$O_{ab} = \int d\mathbf{r} |w_a|^2 |w_b|^2$, where $w_a$ is the Wannier function 
at the link $(r,a)$ and we assume a negligibly small DDI. 
For the horizontal links, by approximating a minimum of the optical lattice 
as a quadratic form 
$m \omega_\mathrm{ho}^2(\alpha^2 x^2 + y^2)/2$, $w_a$ 
can be represented by the harmonic oscillator basis $\Phi_\mathbf{n} (\mathbf{r})$, 
where $\hbar \omega_\mathrm{ho} = 2 \sqrt{E_R( V_\mathrm{A} + 2V_\mathrm{B})}$ 
with the recoil energy $E_R = \hbar^2 k^2/2m$ of the optical lattice and $\alpha 
= \sqrt{(2V_\mathrm{B} - V_\mathrm{A})/(2V_\mathrm{B} + V_\mathrm{A})}$. 
The band index takes $\mathbf{n} = (0,0)$, $(1,0)$, and $(2,0)$ for 
the $s$-, $p$-, and $d$-orbitals. 
For the vertical links, the role of $(x,y)$ is just exchanged by $(y,x)$. 

The conditions in Table~\ref{paratable} read $O_{12} \approx O_{13} \gg O_{15}$. 
Figures~\ref{bandsetup}(b)-(d) represent the parameter domain satisfying 
this condition with respect to $V_\mathrm{A}$ and $V_\mathrm{B}/V_\mathrm{A}$, 
where the amplitudes $V_\mathrm{A}$ and $V_\mathrm{B}$ of the optical lattice are 
precisely tunable parameters.
Because of the characteristics of the potential Eq.~(\ref{suitpot}), we can have 
significant overlap of the Wannier functions even for the high potential height such as 
$V_\mathrm{A,B} = 100 E_R$ for $V_\mathrm{B}/V_\mathrm{A} \geq 0.5$; 
see Appendix~\ref{app2} for more details.  
For the $s$-orbitals the domain is limited to a narrow 
region ($V_\mathrm{B} \sim 0.55 V_\mathrm{A}$) of the parameter space. 
Using the $p$- or $d$- orbitals allows us to get the condition of 
Table~\ref{paratable} more easily 
in the experimentally feasible condition. 
When the excited orbitals are used, we have significant hopping amplitudes $J_{ab}$ 
not only for the NN ($J$) but also the 1st half of 
NNN $(J')$; the 2nd half of NNN ($J''$) 
is small because of higher potential height between the link of group (iii) 
as seen in Fig.~\ref{bandsetup}(a).  

Finally, we admit that, for actual parameter estimation,
one should also try other more realistic Wannier functions such as $\sim x^{c} \exp(-h|x|)$ \cite{He},
although the qualitative feature captured here needs no modifications. 

\subsection{Method B: Using dipolar atoms in a multilayer optical lattice}\label{kunolayer}
\begin{figure}[t]
\centering
\includegraphics[width=1.0\hsize, bb=0 0 567 540]{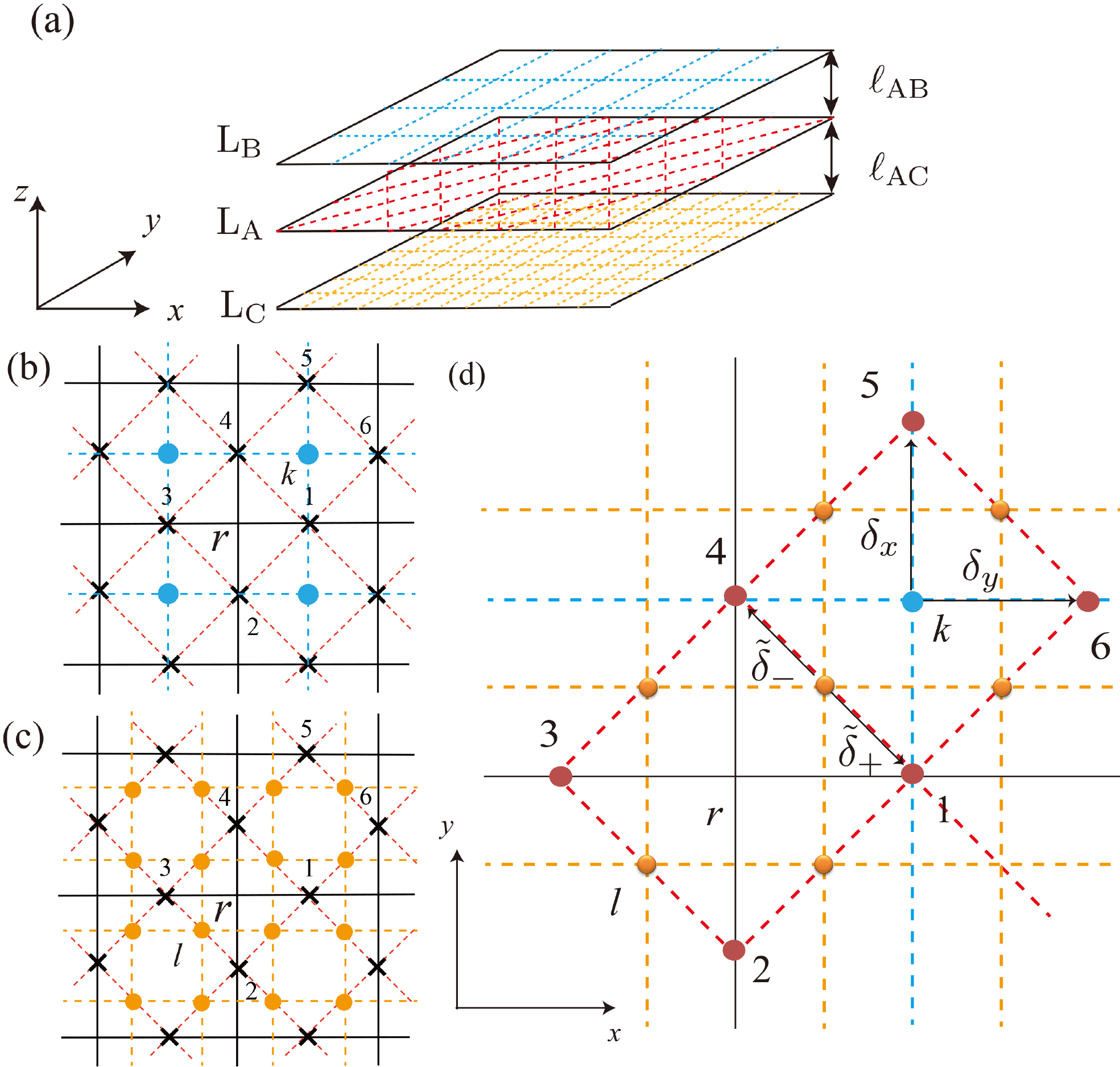}
\caption{(a) The structure of a triple-layer 2D optical lattice. 
The LGT is simulated on the lattice of black solid lines. 
The layers are separated by distance $\ell_\mathrm{AB}$ and $\ell_\mathrm{AC}$.
The panels (b) and (c) show the projective mapping of 
L$_\A$ (red lines) and L$_\B$ (blue lines), and L$_\A$ and L$_\C$ (orange lines), respectively. 
The panel (d) shows the unit cell of the projective mapping of the all layers. 
The site label of L$_\A$ and L$_\C$ are denoted as $k$ and $l$, respectively. 
For the layers L$_\B$ and L$_\A$, we take the NN DDI between 
atoms on the sites 
$k$ in L$_\B$ and $k \pm \delta_{x} (\delta_y)$ in L$_\A$ into account. 
For L$_\C$ and L$_\A$, we take the NN DDI between atoms on the sites
$l$ in L$_\C$ and $l + \tilde{\delta}_{+} (\tilde{\delta}_-)$ in L$_\A$ into account. 
}
\label{OL2}
\end{figure}

The idea for the second method is to introduce new subsidiary 2D lattices
and treat the DDI between atoms in the original 2D lattice and atoms in the
subsidiary lattices by the second-order perturbation theory
to obtain $V_{ab}$ effectively. For illustrative purpose,
we explicitly describe the idea by using a triple-layer 
system consisting of three 2D square optical lattices 
(layer L$_\A$, L$_\B$, L$_\C$) as seen in Fig.~\ref{OL2}(a). 
Here, we neglect the contribution of short-range interaction for the intersite interaction. 
The scheme may be reduced to a double-layer system by approaching the distance between 
two layers, e.g. L$_\A$ and L$_\B$, to zero, which is discussed for realistic parameter 
estimation at the end of Appendix~\ref{app3}. 

The boson system on the layer L$_\A$ (we call them A-bosons) 
is a playground of the (2+1)D U(1) GH model, 
which is sandwiched by B-bosons on L$_\B$ and C-bosons on L$_\C$. 
The B- and C-bosons are trapped in  deep optical lattices with negligible hopping. 
Each layer has different basis vectors of the lattice structure as shown in Fig.~\ref{OL2}(b) and (c). 
Each species of bosons is assumed to have a dipole, 
perpendicular to the plane of the layer. 
By treating the DDI between A-boson and B-boson as a perturbation,
the second-order perturbation theory generates an effective intersite
interaction between the A-bosons. So is the DDI between
A- and C-bosons, which generates another intersite interactions between the A-bosons.
These two kinds of interactions may be tuned to realize 
$V_{ab}$ as given in Table~\ref{paratable}.  
We omit the DDI between the B- and C-bosons because of the large separation. 

Let us focus on Fig.~\ref{OL2}(d). 
When one projects the sites of L$_\B$ onto L$_\A$,
their image locates on the center of each plaquette of the L$_\A$ lattice. 
Similarly, the image of sites of L$_\C$ locates on the middle of NN pairs
of the L$_\A$ sites.
In L$_\A$, the A-bosons 
at different sites have the repulsive DDI. 
Furthermore, the A- and B(C)-bosons are coupled through the NN attractive DDI
given by  
\begin{align}
H_{\A\B} &= V_{\A\B}\sum_{k,\delta} \rho_{A,k+\delta} n_{B,k}, \nonumber  \\ 
H_{\A\C} &= V_{\A\C}\sum_{l,\tilde{\delta}}
\rho_{\A,l+\tilde{\delta}} n_{\C,l},
\label{habc}
\end{align}
where $\rho_{\mathrm{A},k}$ and $n_{\mathrm{B(C)},k}$ are boson densities 
at the site $k$ and $V_{\A\B(\C)} < 0$ is the DDI, which is tunable by controlling the interlayer separation.
Our strategy is to trace out B- and C-bosons to get 
the effective attractive intersite interactions between the A-bosons themselves. 
According to the usual second-order perturbation theory with
$H_{\A\B(\C)}$ as perturbation,
the effective attractive interaction between 
A-bosons may be estimated as $\sim -V^{2}_{\A\B}
\sum_{k,\delta,\delta '} \rho_{\A,k+\delta}\rho_{\A,k+\delta '}$ and
$-V^{2}_{\A\C}\sum_{l,\tilde{\delta}} 
\rho_{\A,l+\tilde{\delta}} \rho_{\A,l-\tilde{\delta}}$.
They are due to density fluctuations of B- and C-bosons, respectively.
The former term contributes a constant to $V_{ab}$ for (a,b)
of the groups (i,iii) of Table~\ref{paratable}, while 
the latter contributes a constant only for the group (i). 
Then one may
fulfill the condition of $V_{ab}$ in Table~\ref{paratable}.
The detailed calculation of the effective interaction and 
the experimental feasibility 
are described in Appendix~\ref{app3}. 
Although there is a small contribution of long-range interaction beyond 
the NNN links due to the power-law tail of $r^3$, this correction may 
suppress the density fluctuation and result in the enhancement of the 
confinement phase.

\section{conclusion and outlook}\label{sums}
In conclusion, realization of the quantum simulator of the U(1) lattice-gauge 
Higgs model provides a significant innovation to tackle unresolved 
problems such as inflation universe, 
being possible to be constructed by the cold-atomic architecture. 
The phase structure of the atomic simulators may be explored 
by the non-equilibrium dynamics, where the electric flux dynamics 
can be observed from the behavior of the density fluctuation. 
We proposed two experimentally feasible schemes (Method A and B) to 
respect the constraint of Gauss's law and locality of the gauge interaction 
in the atomic simulators. 

Many works have been devoted to the dynamical properties of 
phase defects, namely quantized vortices, by analyzing the GP equation \cite{Pethicksimsh}. 
In terms of the gauge theory, these phase defects correspond to the magnetic fluxes. 
Our work focuses on the density fluxes, corresponding to electric fluxes, 
whose dynamics are under constraint by the Gauss's law. 
Such a density flux in the GP model has not been discussed before and 
this point of view could open the door for new avenue of the GP dynamics, 
such as dynamical features of various configurations of an electric flux 
or many fluxes. 
These non-equilibrium dynamics are interesting themselves, although they could 
also give references as a guide not only to the atomic simulator experiments 
but also to the LGT. The dynamical equations can be derived and give some insights 
for various models of the LGT. 

The other problems for the future study includes the clarification 
of the global phase diagram of Eq.~(\ref{KK-zgh}) for the general sets of 
parameters and of how to implement the general terms in Eq.~(\ref{KK-zgh}) 
experimentally. It has been proposed in Ref.~\cite{Kasamatsu} that 
the Higgs coupling ($c_{1i}$-term) in the spatial dimension can be implemented 
by using an idea of Ref.~\cite{Recati}. An idea to generate the spatial 
plaquette ($c_{2ij}$-) term is discussed in Ref.~\cite{Buchler}. 
There is still insufficient discussion on how to combine these schemes toward the 
quantum simulation of the full GH model, which is a subject for future study.
Fine tuning of the intersite density-density interaction is also an important 
task, and we believe that the method in Sec.~\ref{higiorbit} is the most feasible 
scheme in actual experiments. Our method in Sec.~\ref{kunolayer} provides a new scheme 
for tuning the intersite atom-atom interactions, and more elaborated discussion 
using concrete atomic species, optical lattice structures, etc., remains 
to be studied. All of these issues will be reported in future publications. 

\acknowledgements
This work was supported by KAKENHI from JSPS 
(Grant Nos. 26400371, 25220711, 26400246 and 26400412).

\appendix

\section{Phase structure of the U(1) GH model}\label{app1}
\begin{figure}[ht]
\includegraphics[width=1.0\linewidth, bb= 0 0 425 567]{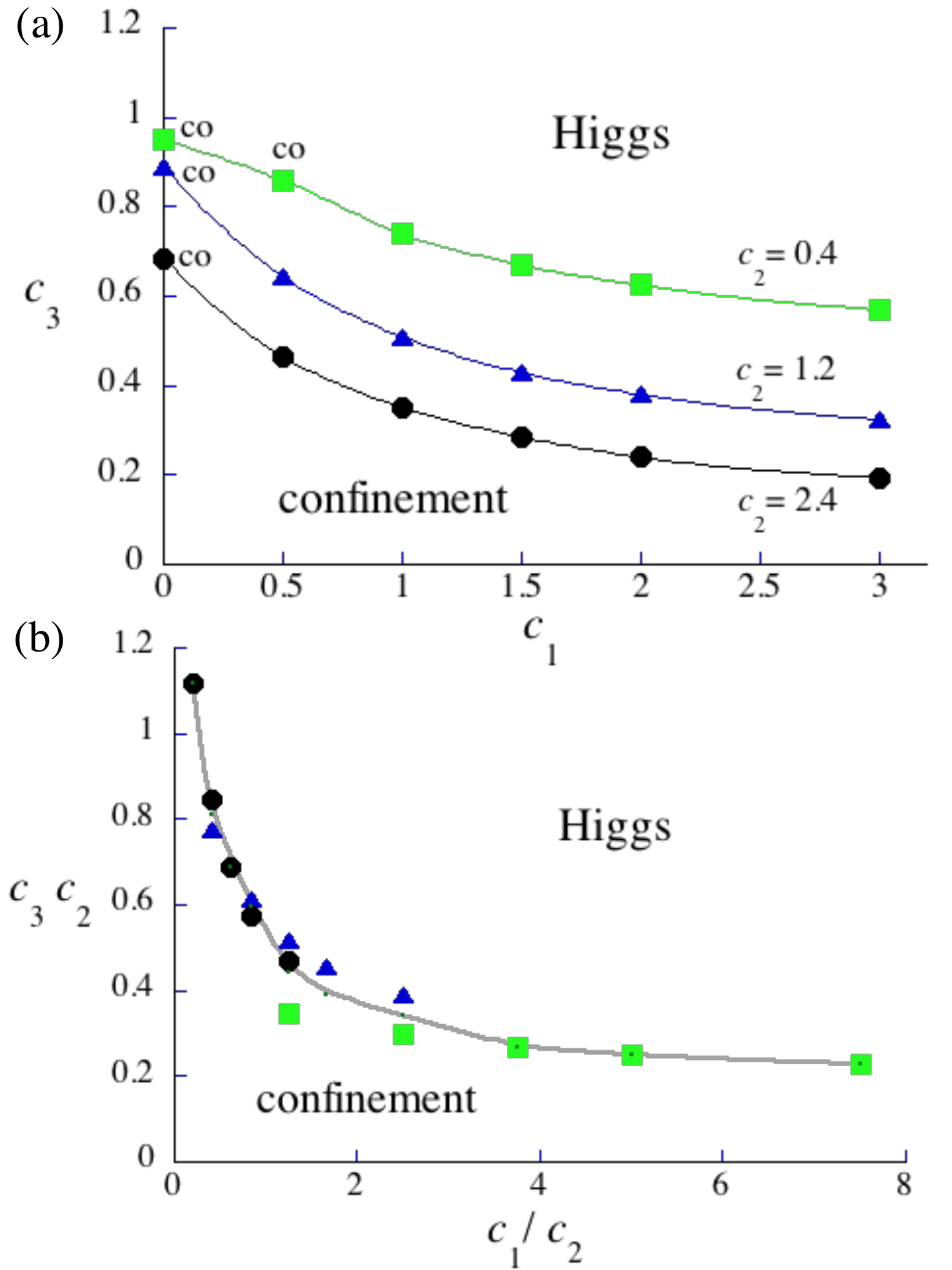}
\caption{
Phase diagram of the (2+1)D U(1) lattice GH model
of Eq.~(\ref{KK-zgh}) with asymmetric couplings $c_{1\mu}$ and $c_{2\mu\nu}$ 
given by Eq.~(\ref{atomicparam}) and $c_4=c_5=0$. 
(a) Three curves connect transition points in the $c_1$-$c_3$ plane
for $c_2=0.4, 1.2, 2.4$ from above,
which separate the Higgs phase (above) and the confinement phase (below).
The transition points are located at the peak of $C$
as a function of $c_3$ for fixed $c_1$. 
They are (i) second-order (no marks) where the peak of $C$ develops 
as the size $L$ increases and $U$ exhibits no hysteresis or
(ii) cross-over or Kosterlitz-Thouless transition (both marked with co)
where the peak does not develop.  (b) The transition points on (a) are arranged in the  
$(c_1/c_2)-(c_3\cdot c_2)$ plane ($\gamma^2V_0'- 2J\rho_0/V_0'$ plane) . They almost sit 
on a single curve (gray line) (See the comment right after Eq.~(\ref{parameterrelation})).}
\label{phasestructure}
\end{figure}

Let us explain the phase structure of the gauge-Higgs model defined
by Eq.~(\ref{KK-zgh}) with asymmetric couplings $c_{1\mu}, c_{2\mu\nu}$ 
given by Eq.~(\ref{atomicparam}).
First, we note that the (2+1)D
version of the standard 4D U(1) Higgs gauge theory \cite{KK-complementarity}, 
which is considered in HEP and has the symmetric couplings
($c_{1\mu}=c_1 \geq 0, c_{2\mu\nu}=c_2 \geq 0, c_{3,4,5} = 0$ in Eq. (\ref{KK-zgh})), 
is always in the confinement phase \cite{KK-Polyakon}, in which the phase $\theta_{x,\mu}$
is unstable by strong fluctuation. In our model, inclusion 
of sufficient $c_3$ in addition to the asymmetric couplings
$c_{1\mu}$ and $c_{2\mu\nu}$ lets
the system enter into the ``Higgs" phase, 
where both $\theta_{x,\mu}$  and $\varphi_x$ are stable [see Fig.~\ref{phasestructure}].

To identify the location of the transitions,
we measure the internal energy $U = \la A \ra$ and 
the specific heat $C =\la A^2\ra-\la A\ra^2$ 
by using the standard Metropolis algorithm in Monte Carlo (MC) simulation 
with the periodic boundary condition
for the cubic lattice of size $V=L^3$ with $L$ up to 40. 
The typical number of sweeps is $100000+10000 \times 10$,
where the first number is for thermalization
and the second one is for measurement. 
The errors of $U$ and $C$ are estimated by the standard deviation over 
10 samples. Acceptance ratios 
in updating variables are controlled to be $0.6\sim 0.8$.

Explicitly, we confine ourselves to the case $c_4=c_5=0$ and
obtain the phase diagram in the $c_1-c_3$ plane
for several values of $c_2$.
The result is presented in Fig.~\ref{phasestructure}. 
There are two phases: the Higgs phase in the large $c_3$ region
(upper region) and the confinement phase in the small $c_3$ region (lower region). 
The confinement-Higgs transition here should correspond
to various phase transitions such as the superconducting transition,
the mass generation in the standard model, and the believed one 
to take place in the early universe \cite{inflation1,inflation2}. 
In contrast to the phase diagram of the (3+1)D model 
for $c_4=c_5=0, c_1=c_3$, and $c_2 \geq 0$ \cite{Kasamatsu}, 
the Coulomb phase is missing due to the low dimensionality.

To understand Fig.~\ref{phasestructure}, let us consider some limiting cases.
First, after choosing the unitary gauge $\phi_x=1$,  
let us consider the limit  $c_1 \to \infty$. Then the  $c_1$ term
makes $\theta_{x,0} = 0$ [mod(2$\pi$)], and  the action becomes     
\be
\hspace{-0.5cm}
A_{\rm c_1=\infty}&=&
c_2\sum_{x}\sum_{i=1}^2\cos(\theta_{x+0,i}-
\theta_{x,i})\nn
\hspace{-0.5cm}
&+& c_3\sum_{x}\Big[
\cos(\theta_{x,1}-\theta_{x,2})+\cos(\theta_{x,1}+\theta_{x+1,2})
\nn
\hspace{-0.5cm}
&+&
\cos(\theta_{x+1,2}-\theta_{x+2,1})+\cos(\theta_{x,2}+\theta_{x+2,1})
\Big],
\label{c1inf}
\ee
up to constant. This is viewed as a 3D XY spin model 
with asymmetric couplings, where
$\theta_{x,i}\ (i=1,2)$ on the link $(x,x+i)$ is the  XY spin angle $\tilde{\theta}_{\tilde{x}}$.
In fact, the $c_3$ term is their NN coupling
in the 12 plane and the $c_2$ term is their NN coupling along the $\mu=0$ axis.
The region of sufficiently large $c_2$ and $c_3$ is the ordered phase of this
XY spins, and corresponds to the Higgs phase with small gauge-field ($\theta_{x,\mu}$) 
fluctuations. 
As a check of Fig. \ref{phasestructure},  
let us consider the case $c_2=c_3$ of Eq.~(\ref{c1inf}), which reduces to the symmetric 3D XY spin model of  
$A_{\rm 3DXY}=c_{\rm XY}\sum_{\tilde{x},\mu}\cos(\tilde{\theta}_{\tilde{x}+\mu}
-\tilde{\theta}_{\tilde{x}})$.
It is known to have a genuine second-order phase transition at $c_{\rm XY}\simeq 0.45$.
Therefore the transition line in Fig.~\ref{phasestructure} (b) should approach 
to $c_2\cdot c_3\to 0.45^2\simeq 0.20 $ as $c_1/c_2\to \infty$ as it shows.

Next, let us consider the case $c_2=0$. Then, each variable $\theta_{x,0}$
appears only through the $c_1$ term without couplings to other variables
(we take the unitary gauge as before).
Then the dynamics is controlled by the $c_3$ term. Again, this term is viewed
as the energy of the XY spins $\theta_{xi}\ (i=1,2)$. However they
 have no coupling along the $\mu=0$ direction, and therefore the system is 
 a collection of decoupled 2D XY spin models.  2D XY spin model
 is known to exhibit Kosterlitz-Thouless (KT) transition which 
 is infinitely continuous.
 Thus, although it is not drawn in Fig.~\ref{phasestructure}, there should be
 added a horizontal line (independent of $c_1$) for $c_2=0$ consisting of 
a collections of KT transitions at around $c_3\sim 0.96$.
We understand that the crossover points appearing in the smaller $c_1$ part
in each curve for three $c_2$ drawn in Fig.~\ref{phasestructure} are the remnants of these KT transitions.
They have a chance to be a genuine KT transition, although we called them
crossover here. 
Another support of this interpretation is to consider the case
$c_1=0$. Then there is no source term for $\theta_{x,0}$ and  
$\theta_{x,0}$ should determine their dynamics only through the $c_2$ term.
Thus, even $\theta_{x,i}$ could be set constant with no fluctuations,
$\theta_{x,0}$ has the NN coupling in each 12 plane.
However, two dimensions is not enough
to stabilize $\theta_{x,0}$. In turn, the $c_2$ term is not enough
to sustain the coupling between $\theta_{x,i}$ along the $\mu=0$ direction.
The dynamics of $\theta_{x,i}$ is essentially from the $c_3$ term,
which is the 2D XY model as explained. Therefore, the transition, if any,
for $c_1=0$ may be a  KT transition. No genuine second-order one is possible.

The last case is $c_2=\infty$. Then $\theta_{x,\mu}$ is frozen to be a 
pure gauge configuration, $\theta_{x,\mu}=\lambda_{x+\mu}-\lambda_x$. 
By plugging this into the $c_1$ and $c_3$ term, we obtain
\be
\hspace{-0.5cm}
&&A_{\rm c_2=\infty}=
c_1\sum_{x}\cos(\lambda_{x+3}-\lambda_x)\nn
\hspace{-0.5cm}
&&+ 2c_3\sum_{x}\Big[
\cos(\lambda_{x+1}-\lambda_{x+2})+\cos(\lambda_{x}-\lambda_{x+1+2})
\Big],
\label{c2inf}
\ee
which belongs again to the class of 3D XY spin models, where
$\lambda_x$ is the XY spin angles on the site $x$.
So we should have a second-order transition at $c_2=\infty$
as long as both $c_1$ and $c_3$ are nonvanishing.
This is consistent with Fig.~\ref{phasestructure}.

Let us finally comment on the transition line of Fig. \ref{phasestructure}(b) and 
the boundaries of Fig.~\ref{Fluxremgp}(a) calculated by dynamical simulation in Sec.~\ref{fluxdyn}.
Their behaviors in the $(c_1/c_2)$-$(c_3\cdot c_2)$ plane are qualitatively consistent 
but different in quantitative comparison.
We understand that there are no inconsistency in these results 
because the two methods, MC and GP, are different in nature:
MC is static  and GP is dynamical, they treat fluctuations in contrasting manners,  
and especially, the dynamical simulations necessarily exhibit various properties of 
the system according to their setup and probes, etc.  
This certainly motivates exact quantum and dynamical simulation of 
the BH model in experiments.

\section{Calculation of overlap integrals}\label{app2}
In this section, we describe the calculation of the overlap integrals discussed in Sec.~III A.
For the horizontal links of the potential $V_\mathrm{OL}$ of Eq.~(\ref{suitpot}), 
the minimum is approximated by the harmonic oscillator $V_{hx} = m \omega_\mathrm{ho}^2(\alpha^2 x^2 + y^2)/2$. 
The basis function of $V_{hx}$ is given by 
\begin{align}
\Phi_\mathbf{n} (\sqrt{\alpha} x,y) = A_\mathbf{n} H_{n_x} \left(\sqrt{\alpha} \frac{x}{a_\mathrm{ho}} \right) 
H_{n_y} \left(\frac{y}{a_\mathrm{ho}} \right) e^{- (\alpha x^2 + y^2)/2 a_\mathrm{ho}^2}, 
\end{align}
where $H_{n}$ is the Hermite polynomial, $A_\mathbf{n}$ the normalization factor, and 
$a_\mathrm{ho} = \sqrt{\hbar/m\omega_\mathrm{ho}}$ the harmonic oscillator length 

The $s$, $p$, and $d$ orbitals for these links correspond to $\mathbf{n}
=(n_x,n_y) = (0,0)$, $(1,0)$, and $(2,0)$. 
As the Wannier function $w_a(\mathbf{r})$ at the link $(r,a)$, we use  
$\Phi_\mathbf{n}$ with $(x, y)$ measured from the center of the link. 
For the vertical links, the minimum is also approximated as 
$V_{hy} = m \omega_\mathrm{ho}^2(x^2 + \alpha^2 y^2)/2$ 
and the basis function is 
$\Phi_\mathbf{n} (x, \sqrt{\alpha} y) = A_n H_{n_x}(x/a_\mathrm{ho}) H_{n_y}(\sqrt{\alpha} y/a_\mathrm{ho}) 
e^{- (x^2 +\alpha y^2)/2a_\mathrm{ho}^2}$. 
The $s$, $p$, and $d$ orbitals for these links correspond to $\mathbf{n}=(n_x,n_y) = (0,0)$, $(0,1)$, and $(0,2)$. 
Then, the Wannier functions $w_a(\mathbf{r})$, relevant to the following calculations, 
are given as follows;
\begin{align}
w_{1} (\mathbf{r}) &= \Phi_\mathbf{n} (\sqrt{\alpha} R_0 (\tilde{x}-1/2) ,
R_0 \tilde{y}), \nn
w_{2} (\mathbf{r}) &= \Phi_\mathbf{n} (R_0 \tilde{x},\sqrt{\alpha} R_0 (\tilde{y}+1/2)),\nn
w_{3} (\mathbf{r}) &= \Phi_\mathbf{n} (\sqrt{\alpha} R_0 (\tilde{x}+1/2) ,R_0 \tilde{y}),\nn
w_{5} (\mathbf{r}) &= \Phi_\mathbf{n} (\sqrt{\alpha} R_0 (\tilde{x}-1/2) , R_0 (\tilde{y}-1), 
\end{align} 
where $R_0$ represents the lattice spacing and we shift the origin of the coordinate 
to $(R_0/2,R_0/2)$ of Fig.~\ref{bandsetup}(a). 
The length scale of the coordinate is normalized by $R_0$ and
the dimensionless coordinates are denoted by putting tildes. 

The intersite interaction strength $V_{ab}$ is
proportional to the overlap integrals
$O_{ab} = \int d\mathbf{r} |w_a|^2 |w_b|^2$.
It is sufficient to  calculate only the 
integrals for the link pairs $(a,b)=(1,2)$, $(1,3)$, and $(1,5)$, because 
$O_{12} = O_{23} = O_{34} =O_{41}$, $O_{13} = O_{24}$, and $O_{15} = O_{26}$
due to the lattice symmetry. 

\begin{figure}[ht]
\centering
\includegraphics[width=0.85\linewidth,bb=0 0 198 425]{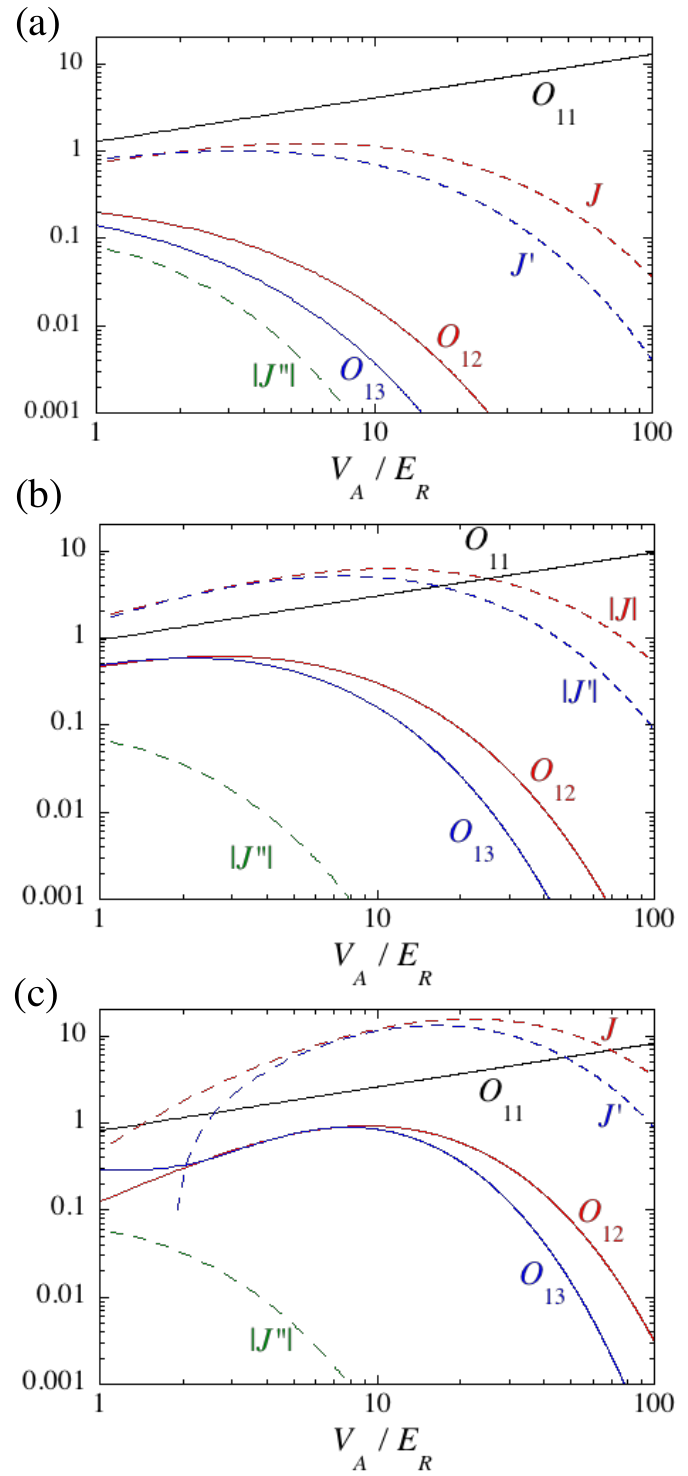}
\caption{Overlap integral $O_{ab}$ and hopping integrals $J$, $J'$, and $J''$ 
as a function of $V_\mathrm{A}/E_R$ for $V_\mathrm{B}/V_\mathrm{A} = 0.6$. 
The panels (a), (b), and (c) corresponds to the cases of $s$-, $p$- , and $d$-wave orbitals. 
Using $O_{ab}$, the density-density interaction is written by $V_{ab}/E_R=(8a/\pi l_z) O_{ab}$ with 
the $s$-wave scattering length $a$ and the typical length scale $l_z$ 
along the $z$-direction. The hopping is also normalized by $E_R$ and its negative value 
is plotted by the absolute value. In the $p$-orbital case, $J_{12}(=J)$ and $J_{13}(=J')$ 
become negative, but there is no frustration because of $J_{23},J_{41} > 0$ and $J_{34},J_{24} < 0$.}
\label{overlap}
\end{figure}

The typical results of the overlap integrals for the three orbitals are shown in Fig.~\ref{overlap} 
for $V_\mathrm{B}/V_\mathrm{A} = 0.6$ as a function of $V_\mathrm{A}$. 
We also show the integral for on-site contribution $O_{11} = \int d\mathbf{r} |w_a|^4 $ and the hopping integrals 
$J=\int d \mathbf{r} w_1 (-\hbar^2 \nabla^2/2m + V_\mathrm{OL}) w_2$, 
$J'=\int d \mathbf{r} w_1 (-\hbar^2 \nabla^2/2m + V_\mathrm{OL}) w_3$, and 
$J''=\int d \mathbf{r} w_1 (-\hbar^2 \nabla^2/2m + V_\mathrm{OL}) w_5$. 
In any case, $O_{11}$ is monotonically increased with $V_\mathrm{A}$, and 
$J''$ and $O_{15}$ (not seen in Fig.~\ref{overlap}) are negligibly small. 
In the case of the $s$-orbital, $O_{12}$ and $O_{13}$ are also monotonically 
decreasing functions, so that the range satisfying $O_{12} \approx O_{13}$ is only 
limited by a narrow range or a point with respect to $V_\mathrm{A}$. 
On the other hand, for the $p$- and $d$-orbitals $O_{12}$ and $O_{13}$ changes 
non-monotonically because of the node structure and the extended amplitude profile 
of the wave functions. This fact extends the range of $O_{12} \approx O_{13}$ 
as seen in Fig.~\ref{overlap}(b) and (c).

Note that the hopping integrals $J$ and $J'$ are of ${\cal O}(1)$ even for $V_\mathrm{A} = 100 E_R$. 
This is because the energy barrier between links of group (i) and (ii) in Table~\ref{vab} is 
the sub-maximum with the height $2V_\mathrm{B}-V_\mathrm{A}$ 
at $(R_0/2,R_0/2)$ in Fig.~\ref{bandsetup}(a). Since, the value of $J(J')$ is bigger than 
that of $O_{12 (13)}$ by two orders of magnitude, one needs to increase considerably 
the $s$-wave scattering length via a Feshbach resonance to get the exact 
Gauss's law constraint, namely, $V_{ab} \gg J$. 

\section{Effective intersite interaction in the triple-layer system of 
Sec.~\ref{kunolayer}} \label{app3}
In this section, we apply the second-order perturbation theory
to the triple-layer system in Sec.~\ref{kunolayer} to derive 
the effective intersite 
interaction of A-bosons, and estimate  the possible values of 
involved parameters to realize $V_{ab}$ of Table~\ref{paratable}.
After that, we briefly explain a double-layer system in which magnitude of the
intersite interaction of A-bosons is controlled in a similar way.
 
We first confine ourselves to the subsystem of the A- and B-bosons
(two layers L$_\A$ and L$_\B$) which has   
the NN DDI, $H_{\A\B}$ of Eq.~(\ref{habc}).
It implies that the B-boson on the site $k$ interacts with the four 
NN A-bosons on the sites 
$k \pm \delta_{x(y)}$ as seen in Fig.~\ref{OL2}(d). 
$V_{\A\B}$ in $H_{\A\B}$ is expressed as 
\be
V_{\A\B} &=& \int d\mathbf{r} d\mathbf{r}' U_{\mathrm{DD}}
(\mathbf{r},\mathbf{r}') |w_{\A}(\mathbf{r})|^2 |w_{\B}(\mathbf{r}')|^2,\nn
U_{\mathrm{DD}}(\textbf{r},\textbf{r}') &=& \frac{C}{4\pi |\textbf{r}-\textbf{r}'|^{3}}
\biggl( 1-\frac{3\ell^{2}_{\mathrm{AB}}}{|\textbf{r}-\textbf{r}'|^{2}}\biggr),
\ee
where $\mathbf{r}(\mathbf{r}')$ is the position of A(B)-boson,
$w_{\A(\B)} (\mathbf{r})$ is their Wannier function,
and $C = \mu_0 \tilde{\mu}_{\A}\tilde{\mu}_{\B}$; 
$\mu_0$ is the magnetic permeability of the vacuum 
and $\tilde{\mu}_{\A(\B)}$ is the magnetic moment of A(B)-atoms. 

We assume that the B-bosons 
of L$_\B$ have a chemical potential $\mu_{\B}$($>0)$, a negligibly small NN hopping
amplitude due to a deep trapping potential, 
an on-site repulsion $U_\B (>0)$, 
and the NN DDI with A-bosons $V_{\A\B}$.
One may forget the DDI between B-bosons, because it is a constant due to 
negligible NN hopping.
Then, 
the Hamiltonian $\hat{H}_\B$ and the partition function $Z_{\B}=
{\rm Tr}\exp(-\beta \hat{H}_\B)$ 
for the subsystem of B-bosons are written by using the B-boson density operator 
$\hat{n}_{k}$ at the site $k$ as 
\begin{align}
\hat{H}_{\B}&= \sum_{k}\Bigl[ -\mu_{\B} \hat{n}_{k} + U_{\B} \hat{n}_{k} 
(\hat{n}_{k}-1)+V_{\A\B}\!\!\sum_{\delta=\pm\delta_{x,y}} \!\! 
\hat{n}_{k}\hat{\rho}_{k+\delta}\Bigr],\nn
Z_{\B} &= \prod_k z_{\B,k},\
z_{\B,k}\!=\!\sum_{n=0}^\infty
\exp \biggl[  -\beta\bigl(E_{n}+nV_{\A\B}\sum_\delta\hat{\rho}_{k+\delta}) 
\bigg],\nn
E_{n}&=-\mu_\B n+U_\B n(n-1).
\end{align}
By assuming $\mu_{\B}, U_{\B} \gg V_{\A\B}$, 
we expand $z_{\B,k}$ up to $O(V_{\A\B}^2)$,
\begin{align}
z_{\B,k} &= \sum_{n=0}^\infty e^{-\beta E_n}
\biggl[1-n W_k+\frac{1}{2} n^2 (W_k)^2\cdots\biggr]\nn
&= F_0 \left[1-\frac{F_1}{F_0}W_k+\frac{1}{2}\frac{F_2}{F_0}(W_k)^2\cdots \right],\nn
W_k&= \beta V_{\A\B}\sum_{\delta}\hat{\rho}_{k+\delta},\hspace{3mm}
F_m\equiv \sum_{n=0}^\infty n^m e^{-\beta E_n}.
\end{align}
Then we have 
\begin{align}
&Z_{\B}=(F_0)^{L^2}\exp\biggl\{\sum_k\Bigl[\la -n \ra_0W_k
+\frac{1}{2}(\Delta n)^2(W_k)^2+\cdots\Bigr]\biggr\}\label{2nd_perturbation},\nn
&\la n^m \ra_0 \equiv  \frac{F_m}{F_0},\hspace{5mm}
(\Delta n)^2\equiv\la n^2\ra_0-\la n \ra_0^2. 
\end{align}
The first-order terms $\propto W_k$ are renormalized to the chemical potential 
of A-bosons
and the second-order terms define the effective density-density interaction
Hamiltonian $\hat{H}_{\A\B\A}$ of A-bosons induced by 
B-boson density fluctuation $(\Delta n)^2$,
\begin{equation}
\hat{H}_{\A\B\A} = -\frac{\beta}{2}(\Delta n)^2 V_{\mathrm{AB}}^2\sum_{k}\sum_{\delta,\delta'}
\hat{\rho}_{k,\delta}\hat{\rho}_{k,\delta'}.
\end{equation}

The DDI between A-atoms and C-atoms can be analyzed in the same way, and we 
obtain another effective density-density 
interaction for the A-bosons, 
$\hat{H}_{\A\C\A}=
-(\beta/2)(\Delta n')^2V^2_{\A\C}\sum_{l,\bar{\delta}, \bar{\delta}'} 
\hat{\rho}_{l+\bar{\delta}}\hat{\rho}_{l+\bar{\delta}'},$
where $(\Delta n')^2$ is obtained by replacing $\mu_\B, U_\B$
by $\mu_\C, U_\C$ in $(\Delta n)^2$.

The sum $\hat{H}_{\A\B\A}+\hat{H}_{\A\C\A}$ contributes 
to the coefficients $V_{ab} $ of the intersite density-density interactions 
for A-bosons as follows;
\begin{eqnarray*}
\hspace{-1.5cm}
&&{\rm For\ NN\ links\ (group \ (i)\ in\ Table~\ref{paratable})},\nn 
\hspace{-1.5cm}
&&V_{ab} = V-\beta (\Delta n)^2V^{2}_{\A\B}
-\beta (\Delta n')^2V^{2}_{\A\C},
\end{eqnarray*}
\be
\hspace{-1cm}
&&{\rm For\ NNN\ links},\nn
\hspace{-1cm}
&&V_{ab}=\left\{
\begin{array}{ll}
V'&\ \mathrm{for\ group \ (ii)} \\
V'-\beta (\Delta n)^2V^{2}_{\A\B} &\ \mathrm{for\ group \ (iii)},
\end{array}\right.
\ee
where $V$ and $V'$ are the direct DDI for NN and NNN link pairs, respectively. 
The condition for $V_{ab}$ in Table~\ref{paratable}
can be established 
by adjusting two inter-layer distances $\ell_{\A\B}$ and $\ell_{\A\C}$ 
and density fluctuations 
$(\Delta n)^2(\mu_\B,V_\B,\beta)$ and
$(\Delta n')^2(\mu_\C,V_\C,\beta)$ as 
\begin{align}
&\beta (\Delta n')^2V^{2}_{\A\C} = V-2V', \nonumber \\
&\beta (\Delta n)^2V^{2}_{\A\B} = V' (=\gamma^{-2}). 
\label{tuningreku}
\end{align}

Let us present some brief account for an example and estimation of the experimental parameters 
that satisfy the tuning relations Eq.(\ref{tuningreku}).
We shall report detailed discussion  on this example and related topics in a future publication.

For bosons loaded in each layer we consider $^{52}$Cr atoms \cite{Griesmaier} 
as A bosons, ${}^{87}$Rb atoms as B bosons, and  ${}^{168}$Er atoms \cite{Aikawa} 
as C bosons. They have the permanent magnetic moments 
$6\mu_{\rm BM}, \mu_{\rm BM}$, and $7\mu_{\rm BM}$
($\mu_{\rm BM}$ is a Bohr magneton), respectively. 
Then we are interested in the effective double-layer system, which is
obtained from the triple-layer system explained above by choosing
$\ell_{\rm AB}=0$.  
The reason for using the double-layer system is to make the intersite interaction 
as large as possible because the magnetic moment of $^{87}$Rb atom is small.

\begin{figure}[t]
\includegraphics[width=1.0\linewidth,bb=0 0  680 283]{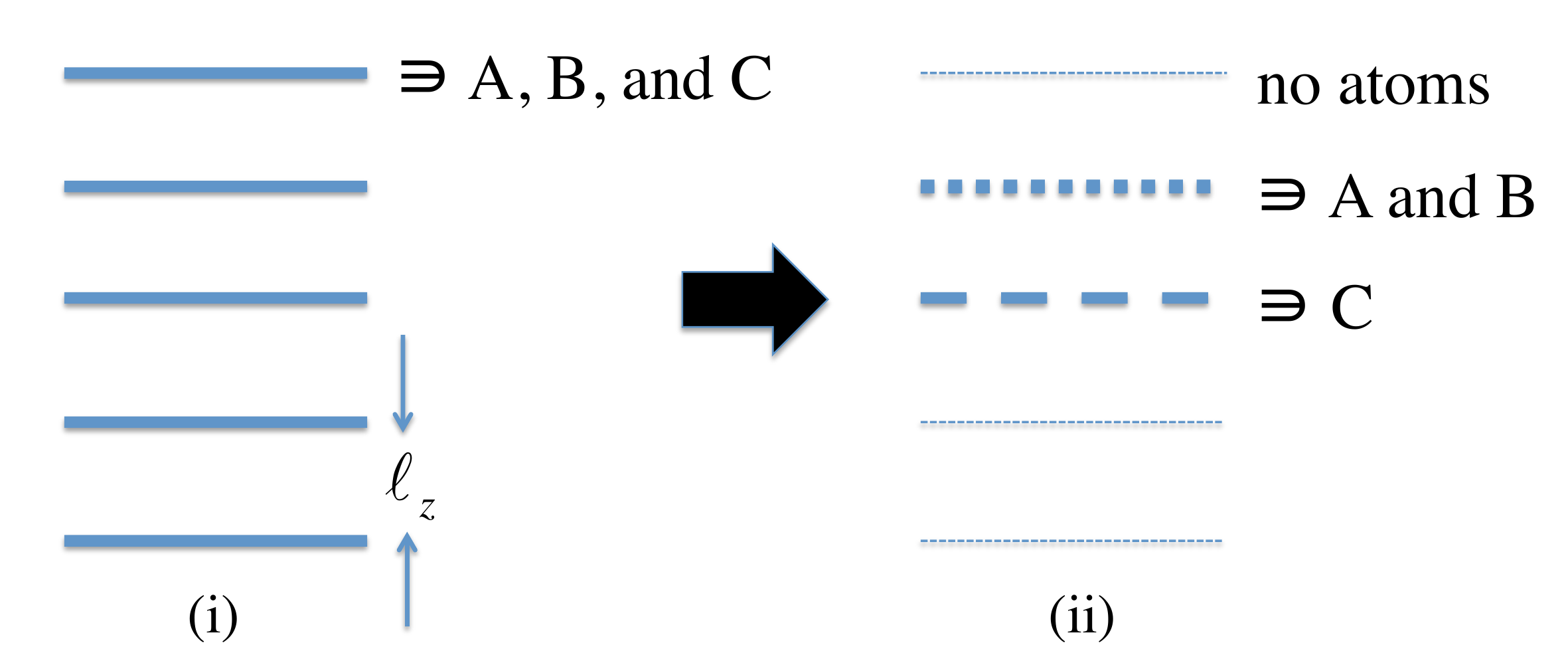}
\caption{Making a double-layer system with three kinds of dipole atoms,
A, B and C.
(i) We prepare a 3D optical lattice. Each $x$-$y$ plane is separated by the distance 
$\ell_z$ and has the 2D lattice structure of  the superposition
of the three layers L$_{\rm A,B,C}$ shown in Fig. 3. Then we supply 
the A,B,C atoms so that every $x-y$ layer is viewed
as the triple-layer system of Fig. 3 with $\ell_{\rm AB}=\ell_{\rm AC}= 0$.
(ii) We blow almost all the atoms off
in such a way that only the A and B atoms in one $x-y$ layer and the C atoms
in the next layer left.
The resulting double-layer system is viewed as the triple-layer system of Fig. 3 with 
$\ell_{\rm AB}=0$ and $\ell_{\rm AC}=\ell_z$.
}
\label{multilayer}
\end{figure}
The method to make such a double-layer system is sketched in Fig.~\ref{multilayer}.
First, one prepares the 3D layer system as shown in Fig.~\ref{multilayer}(i) by emitting
three standing waves with the wavelengths
satisfying 
$2 \lambda_1 = \sqrt{2} \lambda_2 = \lambda_3$
(e.g., $\lambda_{1}=410$nm,  $\lambda_{2}=580$nm 
and $\lambda_{3}=820$nm) in eight appropriate directions 
in the $x$-$y$ plane, each being separated by 45 degrees.  
In addition, we emit another standing-wave laser in the $z$-direction 
with the wavelength $\lambda_{z}$ to establish the 3D structure. 
Because 
$^{52}$Cr,$^{87}$Rb and $^{168}$Er exhibit the specific strong absorptions 
of photon with wave length 425nm, 780nm and 401nm, respectively, 
above standing waves load these atoms to the sites of corresponding 
layer L$_{\A,\B,\C}$ \cite{Bloch_RMP}.
This completes the step (i) in Fig.~\ref{multilayer}.

In the second step (ii) on Fig.~\ref{multilayer}, one needs to
remove almost all the atoms except for those in two adjacent $x$-$y$  layers.
This can be experimentally realized by using the technique of a position-dependent 
microwave transfer in a magnetic field gradient perpendicular 
to the layers \cite{Bloch_nat} successively.    
This achieve to make an effective double-layer system with $\ell_{\rm AB}=0,
\ell_{\rm AC}=\ell_z$.  

Finally, let us estimate the parameters to satisfy the tuning relation Eq.~(\ref{tuningreku}). 
By making a straightforward calculation using DDI,
we find that the following is a typical  example of parameters :
\begin{align}
&V \sim \frac{36\mu_{0} \mu_{\rm BM}^2}{4\pi \lambda^{3}_{2}}, \hspace{5mm} 
\ell_{\rm AC}=\ell_{z} \sim 580 \ [\mathrm{nm}],\nn
&\beta \sim \frac{1}{2V}, \hspace{5mm} U_\B \sim 0.3V,\hspace{5mm} 
\mu_\B\sim 2.5V,\nn
& U_\C\sim 1.3V ,\hspace{5mm} \mu_\C \sim 2V.
\label{ddiestimation2}
\end{align}
The average densities per site  are $\sim 560$ for B bosons 
and $\sim 8$ for C bosons. 
The ratio $|V_{\rm AB(C)}/\mu_{\rm B(C)}|$ is $\sim 0.192$($\sim 0.585$), 
which seems to validate the perturbation theory.



\begin{thebibliography}{99}

\bibitem{book}
M. Lewenstein, A. Sanpera, and V. Ahufinger, 
\textit{Ultracold Atoms in Optical Lattices: Simulating Quantum Many-body Systems} 
(Oxford: Oxford University Press, 2012).
%
\bibitem{KK-wilson}
K. Wilson, 
Phys. Rev. D \textbf{10}, 2445 (1974). 
%
\bibitem{KK-kogut}
J. B. Kogut, 
Rev. Mod. Phys. \textbf{51}, 659 (1979).
%
\bibitem{Wiese}
U. -J. Wiese, 
Annalen der Physik \textbf{525}, 777 (2013).
%
\bibitem{Zohar2}
E. Zohar, J. I. Cirac, and B. Reznich, 
Phys. Rev. Lett. \textbf{109}, 125302 (2012).
%
\bibitem{Tagliacozzo1}
L. Tagliacozzo, A. Celi, A. Zamora, and M. Lewenstein, 
Ann. Phys. \textbf{330}, 160 (2013).
%
\bibitem{Banerjee1}
D. Banerjee, M. Dalmonte, M. M\"uller, E. Rico, P. Stebler, U.-J. Wiese, and P. Zoller, 
Phys. Rev. Lett. \textbf{109}, 175302 (2012).
%
\bibitem{Zohar3}
E. Zohar, J. I. Cirac, and B. Reznich, 
Phys. Rev. Lett. \textbf{110}, 055302 (2013).
%
\bibitem{Zohar4}
E. Zohar, J. I. Cirac, and B. Reznich, 
Phys. Rev. Lett. \textbf{110}, 125304 (2013). 
%
\bibitem{Banerjee2}
D. Banerjee, M. B\"ogli, M. Dalmonte, E. Rico, P. Stebler, 
U.-J. Wiese, and P. Zoller, 
Phys. Rev. Lett. \textbf{110},  125303 (2013). 
%
\bibitem{Tagliacozzo2}
L. Tagliacozzo, A. Celi, P. Orland, M. W. Mitchell, and M. Lewenstein, 
Nat. Commun. \textbf{4}, 2615 (2013). 
%
\bibitem{Horn}
D. Horn, 
Phys. Lett. B \textbf{100}, 149 (1981).
%
\bibitem{Orland}
P. Orland and D. Rohrlich, 
Nucl. Phys. B \textbf{338}, 647 (1990).
%
\bibitem{Chandrasekharan}
S. Chandrasekharan and U.-J Wiese, 
Nucl. Phys. B \textbf{492}, 455 (1997).
%
\bibitem{KK-ks}
J. Kogut and L. Susskind, 
Phys. Rev. D \textbf{11}, 395 (1975).
%
\bibitem{Zohar5}
E. Zohar, J. I. Cirac, and B. Reznik
Phys. Rev. A \textbf{88}, 023617 (2013).
%
\bibitem{Zohar1}
E. Zohar and B. Reznik, 
Phys. Rev. Lett. \textbf{107}, 275301 (2011). 
%
\bibitem{KK-Tewari}
S. Tewari, V. W. Scarola, T. Senthil, and S. Das Sarma, 
Phys. Rev. Lett. \textbf{97}, 200401 (2006).
%
\bibitem{Kasamatsu}
K. Kasamatsu, I. Ichinose, and T. Matsui, 
Phys. Rev. Lett. \textbf{111}, 115303 (2013). 
%
\bibitem{KK-complementarity}
E. Fradkin and S. H. Shenker, 
Phys. Rev. D \textbf{19}, 3682 (1979).
%
%
\bibitem{btj}
K. Aoki, K. Sakakibara, I. Ichinose and  T. Matsui, 
Phys. Rev. B \textbf{80}, 144510 (2009). 
%
\bibitem{inflation1}
A. H. Guth, 
Phys. Rev. D \textbf{23}, 347  (1981).
%
\bibitem{inflation2}
E. Kolb and M. Turner, 
 \textit{The Early Universe}, (Boulder, CO: Westview Press, 1994).
%
%
%
%
%

%

\bibitem{mielke}
J. Motruk and A. Mielke,
J. Phys. A \textbf{45} (2012) 225206.

\bibitem{dipolerev}
T. Lahaye, C. Menotti, L. Santos, M. Lewenstein, and T. Pfau, 
Rep. Prog. Phys. \textbf{72}, 126401 (2009).

\bibitem{Scarola}
V. M. Scarola and S. Das Sarma, 
Phys. Rev. Lett. \textbf{95}, 033003 (2005). 

\bibitem{Kevbook}
P. G. Kevrekidis, 
\textit{The Discrete Nonlinear Schr\"{o}dinger Equation}, 
(Berlin: Springer, 2009).

\bibitem{Reinhard}
A. Reinhard, J. Riou, L. A. Zundel, D. S. Weiss, S. Li, A. M. Rey, and R. Hipolito,
Phys. Rev. Lett. \textbf{110}, 033001 (2013). 

\bibitem{Wenzel}
S. Wenzel, E. Bittner, W. Janke, A. M. J. Schakel, and A. Schiller,
Phys. Rev. Lett. \textbf{95}, 051601 (2005). 

\bibitem{Yan}
B. Yan, S. A. Moses,	B. Gadway, J. P. Covey, K. R. A. Hazzard,	A. M. Rey, D. S. Jin, and J. Ye,
Nature(London) \textbf{501}, 521 (2013). 

\bibitem{Paz}
A. de Paz, A. Sharma, A. Chotia, E. Mar\'echal, J. H. Huckans, P. Pedri, L. Santos, O. Gorceix, L. Vernac, and B. Laburthe-Tolra,
Phys. Rev. Lett. \textbf{111}, 185305 (2013). 

\bibitem{Gorshkov}
A. V. Gorshkov, S. R. Manmana, G. Chen, E. Demler, M. D. Lukin, and A. M. Rey, 
Phys. Rev. A \textbf{84}, 033619 (2011). 

\bibitem{Moessner}
G. M\"{o}ller and R. Moessner, 
Phys. Rev. Lett. \textbf{96}, 237202 (2006). 

\bibitem{Glaetzle}
A. W. Glaetzle, M. Dalmonte, R. Nath, I. Rousochatzakis, R. Moessner, and P. Zoller, 
Phys. Rev. X \textbf{4}, 041037 (2014). 

\bibitem{Macia}
A. Macia, G. E. Astrakharchik, F. Mazzanti, S. Giorgini, J. Boronat,
Phys. Rev. A \textbf{90}, 043623 (2014). 

\bibitem{Muller}
T. M\"{u}ller, S. F\"{o}lling, A. Widera, and I. Bloch, 
Phys. Rev. Lett. \textbf{99}, 200405 (2007). 

\bibitem{Wirth}
G. Wirth, M. \"{O}lschl\"{a}ger, and A. Hemmerich, 
Nat. Phys. \textbf{7}, 147 (2011). 

\bibitem{Olschlager}
M. \"{O}lschl\"{a}ger, G. Wirth, and A. Hemmerich, 
Phys. Rev. Lett. \textbf{106}, 015302 (2011). 

\bibitem{He}
L. He and D. Vanderbilt
Phys. Rev. Lett. \textbf{86}, 5341 (2001). 

\bibitem{Pethicksimsh}
C. J. Pethick and H. Smith
\textit{Bose-Einstein Condensation in Dilute Gases, 2nd ed.}, 
(Cambridge Univ. Press, Cambridge, 2008). 

\bibitem{Recati}
A. Recati, P. O. Fedichev, W. Zwerger, J. von Delft, and P. Zoller, 
Phys. Rev. Lett. \textbf{94}, 040404 (2005).

\bibitem{Buchler}
H. P. B\"uchler, M. Hermele, S. D. Huber, M. P. A. Fisher, and P. Zoller, 
Phys. Rev. Lett. \textbf{95}, 040402 (2005).
%
\bibitem{KK-Polyakon}
A. M. Polyakov, 
Phys. Lett. B \textbf{59}, 82 (1975).

\bibitem{Griesmaier}
A. Griesmaier, J. Werner, S. Hensler, J. Stuhler, and T. Pfau, 
Phys. Rev. Lett. \textbf{94}, 160401 (2005).

\bibitem{Aikawa}
K. Aikawa, A. Frisch, M. Mark, S. Baier, A. Rietzler, R. Grimm, and F. Ferlaino, 
Phys. Rev. Lett. \textbf{108}, 210401 (2012).

\bibitem{Bloch_RMP}
I. Bloch, J. Dalibard, and W. Zwerger, 
Rev. Mod. Phys. \textbf{80}, 885 (2008).

\bibitem{Bloch_nat}
J. F. Sherson, C. Weitenberg, M. Endres, M. Cheneau, I. Bloch and S. Kuhr, 
Nature. \textbf{467}, 68-72 (2010).

\end{thebibliography}
\end{document}